\documentclass[10pt,journal,letterpaper]{IEEEtran}

\usepackage{cite}
\usepackage[utf8]{inputenc}
\usepackage[T1]{fontenc}
\usepackage{amsmath,amssymb,amsfonts}
\usepackage{booktabs}
\usepackage{microtype}
\usepackage{graphicx}
\usepackage[caption=false,font=footnotesize]{subfig}
\usepackage{siunitx}
\usepackage{threeparttable}
\usepackage{multirow}
\usepackage{xcolor}
\usepackage{tabularx}
\usepackage{array}
\usepackage{url}
\usepackage{hyperref}
\hypersetup{hidelinks}

\newcolumntype{Y}{>{\raggedright\arraybackslash}X}
\newlength{\pretrainvizheight}
\definecolor{markGreen}{HTML}{139C78}
\definecolor{markRed}{HTML}{D75A5A}
\definecolor{bgYellow}{HTML}{D9B268}
\definecolor{bgBlue}{HTML}{8EB9D3}
\definecolor{bgPink}{HTML}{D9959D}

\title{Understanding Self-Supervised Representation Learning for Encrypted Traffic}

\author{Zongxi~Liu, Shuai~Mao, Jie~Chuai, and Hongyang~Du%
\thanks{Zongxi Liu and Hongyang Du are with the Department of Electrical and Computer Engineering, The University of Hong Kong, Hong Kong SAR, China (e-mail: duhy@hku.hk).}%
\thanks{Shuai Mao and Jie Chuai are with Huawei Noah's Ark Lab, Hong Kong SAR, China.}}

\begin{document}
\maketitle
\raggedbottom


\begin{abstract}
    Self-supervised learning (SSL) has been adopted for encrypted traffic classification to learn representations from unlabeled traces and reduce reliance on manual feature engineering. However, recent studies show that data leakage can inflate reported performance and that pretrained models can underperform shallow classifiers once leakage is controlled. 
    These findings raise fundamental questions about what information SSL can learn from encrypted traffic and how different pretraining mechanisms exploit this information. In this paper, we systematically investigate these questions by examining the sources of learnable information, the effectiveness of contextual pretraining, alternative self-supervision mechanisms, and the limits of representation learning. 
    We first find that visible protocol fields retain contextual dependencies that are strongest between adjacent bytes, while payload interventions establish the classification value of retained plaintext structure.
    Guided by these insights, we investigate byte-level contextual learning through ByteBERT and TrafficDCNet, showing that aligning prediction granularity with local dependencies produces more discriminative representations than existing contextual pretraining approaches. 
    We further examine cross-view contrastive learning through AlignTraffic and find that complementary supervision improves representation quality beyond contextual prediction alone. 
    Across four public classification tasks, AlignTraffic outperforms shallow learning by 5.97 to 30.70 percentage points in Macro-F1, with additional evaluations on industrial traffic and geographic transfer demonstrating its effectiveness in broader settings. Finally, evaluations on Tor traffic and individual packets reveal limitations when application information is concealed or unevenly distributed. These findings provide a systematic understanding of how observable information and self-supervision mechanisms shape the effectiveness and scope of representation learning for encrypted traffic.
    
\end{abstract}


\section{Introduction}
Traffic classification is an important task in network measurement and security analysis, as it aims to understand system behavior, application usage, and security events by identifying applications or services from their communications. Early systems relied on transport ports and deep packet inspection (DPI), which recognizes signatures in packet contents~\cite{haffner2005acas}. However, the widespread adoption of Transport Layer Security (TLS) has made much of that content opaque, reducing the information available to payload inspection~\cite{rescorla2018transport}. In response, classifiers instead use observable statistical properties of traffic through feature engineering without parsing the encrypted payload~\cite{draper2016characterization,taylor2016appscanner}. These task-specific approaches relying on feature engineering nevertheless offer limited transferability across classification settings.

More recently, advances in self-supervised pretraining in natural language processing (NLP) and computer vision (CV) have motivated learning traffic representations from unlabeled traces to reduce reliance on feature engineering. ET-BERT adapts BERT-style masked language modeling (MLM) to overlapping two-byte units~\cite{10.1145/3485447.3512217}. Subsequent models, including YaTC and NetMamba, explore different input granularities and contextual pretraining objectives~\cite{zhao2023yet,wang2024netmamba}. However, recent analyses question whether encrypted traffic provides the contextual dependencies assumed by these methods. Zhao et al.~\cite{10.1145/3718958.3750498} report weak pretrained representations and show that data leakage from packet-level splitting can inflate downstream classification accuracy. Once these shortcuts are controlled, the advantage of contextual pretraining over shallow classifiers becomes much less evident~\cite{wang2026bias}. These findings motivate us to revisit the basis and effectiveness of pretraining for encrypted traffic.

From a cryptographic perspective, ciphertext produced by secure encryption is expected to offer little exploitable byte-level context~\cite{rogaway2004nonce}. Yet observed TLS traces also contain unencrypted protocol metadata, including record headers and parts of the handshake. These visible fields may retain dependencies that support pretraining even when ciphertext does not. Understanding the effectiveness of encrypted-traffic pretraining therefore requires considering both the remaining observable structure and effective signal sources of self-supervision. We organize this investigation around four questions.

\textbf{Q1. What information remains learnable in encrypted traffic?} Observed encrypted traffic contains both protocol-visible plaintext and ciphertext. Whether the visible fields retain contextual dependencies useful for pretraining is central to establishing what can be learned.

\textbf{Q2. How can contextual pretraining better exploit the available dependencies?} If useful dependencies exist, their range, strength, and distribution determine how they can support prediction. These characteristics should guide the choice of input units and learning objectives.

\textbf{Q3. Can alternative pretraining paradigms be effective for encrypted traffic?} When contextual dependencies are absent or weak, other sources of supervision like contrastive learning may still support representation learning.

\textbf{Q4. What are the limits of pretraining for encrypted traffic?} Traffic settings differ in the application information they expose, and proxy encapsulation can further conceal that information. Whether current pretraining methods remain effective across these settings determines their practical scope.

Our findings are as follows.
\begin{itemize}
    \item Protocol-visible fields retain useful local contextual dependencies. Their byte dependence is substantially stronger than that measured in ciphertext, and the retained payload structure carries classification-relevant information.
    \item Byte dependencies in protocol-visible regions are strongest between adjacent bytes, yet existing contextual pretraining schemes do not fully exploit this local structure. Experiments with ByteBERT and TrafficDCNet show that matching prediction granularity to these dependencies yields more discriminative representations and improves performance over the existing schemes.
    \item Cross-view contrastive learning complements contextual prediction. AlignTraffic improves frozen representations over both constituent encoders and exceeds shallow learning in fine-tuned Macro-F1 on the main public tasks. Industrial classification and geographic transfer provide further evidence of its utility, although these gains come with additional computational overhead.
    \item The effectiveness of pretraining depends on the available application information and the classification unit. Some encapsulated-traffic and individual-packet settings remain challenging for the evaluated representations.
\end{itemize}

\section{Preliminary}\label{sec:setup}
\subsection{Datasets and Preprocessing}
\begin{figure}[!tb]
    \centering
    \includegraphics[width=\linewidth]{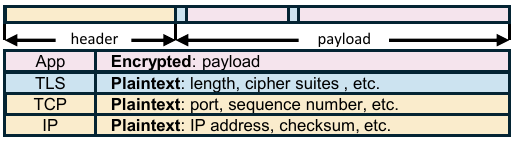}
    \caption{Protocol stack and observable fields in TLS-encrypted traffic. The payload includes visible TLS metadata and encrypted content.}
    \label{fig:stack}
\end{figure}

TLS is the most widely deployed protocol for securing Internet communications. As shown in Fig.~\ref{fig:stack}, a classic TLS packet contains IP and TCP headers followed by a transport payload carrying TLS records. The outer headers expose addressing and transport metadata. Within the payload, TLS record headers reveal framing information, including record length, and unencrypted handshake messages expose selected protocol parameters.
The retained payload therefore combines visible protocol fields with ciphertext.

We use the ISCX suite and CSTNET to study encrypted traffic from different collection environments. We reorganize ISCX traces by access setting into ISCX-Direct, which uses a direct connection, and ISCX-VPN, which uses a VPN proxy~\cite{draper2016characterization}. CSTNET contains TLS~1.3 traffic collected on the China Science and Technology Network~\cite{10.1145/3485447.3512217}.

The original ISCX traces also include plaintext and non-target traffic. Following the protocol-based preprocessing of Zhao et al.~\cite{10.1145/3718958.3750498}, we remove plaintext flows using protocol types decoded by \texttt{tshark}. We then measure the zero-order entropy of the retained payloads. Table~\ref{tab:dataset_intro} reports post-filtering values of 7.97 and 7.95 bits for ISCX-Direct and ISCX-VPN, respectively, close to the 8-bit maximum for a byte.

\begin{table}[!tb]
    \centering
    \caption{Dataset statistics and payload entropy before and after preprocessing.}
    \label{tab:dataset_intro}
    \begin{threeparttable}
        \small
        \setlength{\tabcolsep}{2.5pt}
        \begin{tabular}{@{}lrrrr@{}}
            \toprule
                        & \multicolumn{2}{c}{Packets} & \multicolumn{2}{c}{$H_0(\mathcal{S})$}                  \\
            \cmidrule(lr){2-3}\cmidrule(lr){4-5}
            Dataset     & Before                      & After                                  & Before & After \\
            \midrule
            ISCX-Direct & \num{15715358}              & \num{2776290}                          & 5.70   & 7.97  \\
            ISCX-VPN    & \num{2816966}               & \num{321414}                           & 7.08   & 7.95  \\
            CSTNET      & \num{12049643}              & --                                     & 7.97   & --    \\
            \bottomrule
        \end{tabular}
    \end{threeparttable}
\end{table}

\subsection{Tasks, Inputs, and Evaluation}
We consider two classification tasks. Flow-level classification assigns an application label to a bidirectional TCP flow using its first five retained packets in temporal order. Packet-level classification predicts the parent flow's label from one packet. In both cases, we partition flows before extracting samples, keeping all packets from the same flow in one partition~\cite{10.1145/3718958.3750498}. We follow the CSTNET split protocol of Zhao et al. and use a category-wise 50\%/50\% flow split for the reorganized ISCX data. Model selection uses three-fold validation within the training partition. CSTNET and ISCX-Direct support flow-level evaluation, while ISCX-VPN, with only 966 independent retained flows, is evaluated at packet level. We also consider packet-level ISCX-Direct and CSTNET.

\begin{figure}[!tb]
    \centering
    \includegraphics[width=\linewidth]{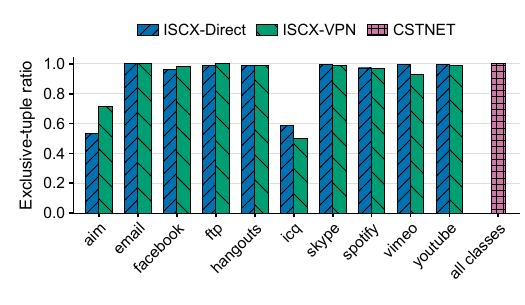}
    \caption{Application-exclusive 5-tuple ratios. ISCX-Direct and ISCX-VPN show per-application ratios. CSTNET shows a weighted mean across its applications, with weights proportional to their numbers of distinct 5-tuples.}
    \label{fig:header_analysis}
\end{figure}

Source and destination identifiers can provide collection-specific shortcuts. For each application, we measure the fraction of its distinct 5-tuples that occur in no other application. Fig.~\ref{fig:header_analysis} shows that many tuples are application-exclusive. In unfrozen flow-level CSTNET classification, retaining these identifiers raises ET-BERT's Macro-F1 from 60.04\% to 85.23\% and YaTC's from 74.42\% to 92.92\%. We therefore mask IP addresses, transport ports, and the remaining 5-tuple fields in all main public comparisons.

We report accuracy (Acc) and Macro-F1. Acc measures the overall fraction of correct predictions, while Macro-F1 averages class-specific F1 scores with equal weight and therefore captures recognition across imbalanced classes. We report both frozen and unfrozen results. In the frozen setting, the pretrained encoder is fixed and only the classifier is trained, which directly reflects the quality of the learned representation. In the unfrozen setting, the entire model is fine-tuned to reflect end-to-end performance.

Our contextual baselines are ET-BERT, YaTC, and NetMamba~\cite{10.1145/3485447.3512217,zhao2023yet,wang2024netmamba}. We also compare with Pcap-Encoder and shallow classifiers, both of which have been reported to outperform contextual pretraining baselines under flow-disjoint evaluation~\cite{10.1145/3718958.3750498,breiman2001random,chen2016xgboost,ke2017lightgbm}. Our models comprise two byte-level encoders, ByteBERT and TrafficDCNet, and their alignment framework, AlignTraffic. We introduce their designs in Sections~\ref{sec:q2} and~\ref{sec:q3}. Public baselines use released implementations or checkpoints. Appendix~\ref{app:hyper}and~\ref{app:shallow} provides reproducibility information.

\section{Learning from Protocol-Visible Structure}\label{sec:q1}
\subsection{Entropy and Mutual Information}
We first examine byte-level dependence in the retained payloads. For a sequence $\mathcal{S}=(s_1,\ldots,s_{N_s})$ over the byte alphabet $\mathcal{X}=\{0,\ldots,255\}$, let $P(x)$ denote the empirical byte distribution. At lag $k\in\{1,\ldots,N_s-1\}$, let $P_k(x,y)$ be the empirical joint distribution of pairs $(s_i,s_{i+k})$ with $1\leq i\leq N_s-k$. Its marginals are $P_k(x)$ and $P_k(y)$, with $P_k(y\mid x)=P_k(x,y)/P_k(x)$.

We compute marginal entropy, conditional entropy, and mutual information as
\begin{equation}
    \begin{aligned}
        H_0(\mathcal{S}) & =-\sum_{x\in\mathcal{X}}P(x)\log_2 P(x),                              \\
        H_k(\mathcal{S}) & =-\sum_{x,y\in\mathcal{X}}P_k(x,y)\log_2P_k(y\mid x),                 \\
        I_k(\mathcal{S}) & =\sum_{x,y\in\mathcal{X}}P_k(x,y)\log_2\frac{P_k(x,y)}{P_k(x)P_k(y)}.
    \end{aligned}
    \label{eq:payload_dependency}
\end{equation}
$H_0$ measures marginal byte uncertainty and is at most 8 bits. $H_k$ measures the uncertainty in the later byte given the earlier byte, while $I_k$ measures their shared information. We compute these statistics over the full retained payload and additionally estimate $I_k$ within its protocol-visible and ciphertext regions.

\begin{figure}[!tb]
    \centering
    \includegraphics[width=\columnwidth]{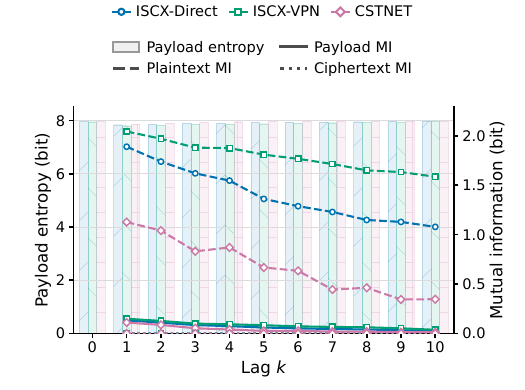}
    \caption{Payload entropy and mutual information across byte lags. Bars show marginal entropy $H_0$ at $k=0$ and conditional entropy $H_k$ at positive lags. Solid, dashed, and dotted curves show mutual information for the full payload, protocol-visible plaintext, and ciphertext, respectively.}
    \label{fig:payload_analysis}
\end{figure}

Fig.~\ref{fig:payload_analysis} shows a consistent pattern across ISCX-Direct, ISCX-VPN, and CSTNET. Although full-payload conditional entropy remains close to 8 bits, nonzero mutual information reveals residual byte dependence that peaks between adjacent bytes and generally decreases with lag. Region-specific analysis indicates that this dependence comes primarily from protocol-visible plaintext, which exhibits substantially stronger dependence and follows the same pattern across lags. Ciphertext mutual information remains close to zero across the measured lags, consistent with its expected randomness. The local dependencies retained in visible protocol fields therefore provide a basis for contextual pretraining.

\subsection{Payload Interventions on CSTNET}
\begin{figure}[!tb]
    \centering
    \includegraphics[width=\columnwidth]{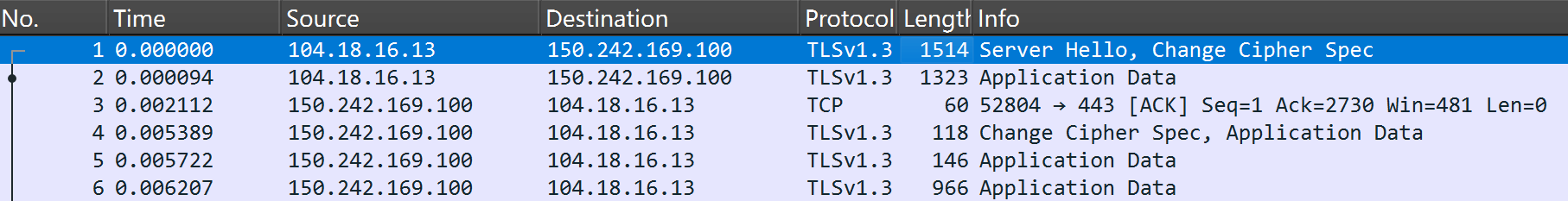}
    \caption{An example of the retained five-packet prefix in CSTNET. The first retained packet contains server-side TLS handshake information, while the following packets primarily contain TLS-encrypted payload.}
    \label{fig:wireshark}
\end{figure}

To further assess the classification value of the visible protocol fields, we conduct controlled payload interventions on CSTNET. CSTNET removes the TCP handshake and the initial client TLS ClientHello. Its first retained packet can still contain visible server-side handshake information, while later packets primarily carry encrypted payload. Fig.~\ref{fig:wireshark} illustrates a five-packet prefix with this structure.

We therefore examine whether the model can exploit the plaintext protocol information retained in the payload for application classification. Using AlignTraffic as a classification probe, we compare the original input with versions that shuffle only the first packet's payload or only the remaining four packets' payloads. We also shuffle all payload bytes or replace them with random values as two additional controls.

Table~\ref{tab:cstnet-process} shows that shuffling only the remaining four packets yields 89.04\% accuracy, close to the original 90.35\%. Shuffling the first retained packet instead reduces accuracy to 61.14\%, comparable to shuffling the entire payload at 63.28\% or randomizing it at 60.63\%. The classification-relevant payload structure is thus concentrated in the first retained packet. Together with the region-specific dependence measurements, these results support the role of visible protocol fields in application classification.

\begin{table}[!tb]
    \centering
    \caption{Unfrozen flow-level classification on CSTNET under payload interventions using AlignTraffic. Acc is in percentage.}
    \label{tab:cstnet-process}
    {\small\begin{tabular}{@{} lc @{}}
            \toprule
            Payload intervention              & Acc            \\
            \midrule
            Original                          & \textbf{90.35} \\
            Shuffle payload                   & 63.28          \\
            Random payload                    & 60.63          \\
            Shuffle only 1st packet payload   & 61.14          \\
            Shuffle except 1st packet payload & 89.04          \\
            \bottomrule
        \end{tabular}}
\end{table}

\section{Exploiting Local Dependencies with Byte-Level Models}\label{sec:q2}
\subsection{Prediction Granularity and Local Context}
Section~\ref{sec:q1} shows that protocol-visible plaintext in encrypted traffic retains contextual dependencies and provides useful information for application classification. However, Zhao et al.~\cite{10.1145/3718958.3750498} report that shallow classifiers can match or outperform existing pretrained models after controlling for data leakage. This motivates us to examine whether current input representations and pretraining objectives allow models to exploit the available protocol structure effectively.

We notice that Fig.~\ref{fig:payload_analysis} shows that byte dependence in protocol-visible regions is strongest between adjacent bytes and generally decreases with lag. To exploit this local structure, the input units and masking policy should preserve informative neighboring bytes as context.

However, ET-BERT's tokenizer encodes overlapping two-byte units~\cite{10.1145/3485447.3512217}. For bytes $a,b,c,d$, the units are $(a,b)$, $(b,c)$, and $(c,d)$. When the middle unit is masked but its neighbors remain visible, both target bytes are still present in the input. Token recovery can therefore exploit these overlapping copies, weakening the need to learn dependencies between distinct original bytes.

YaTC masks whole patches and minimizes mean squared error (MSE)~\cite{zhao2023yet}. Adjacent bytes within a masked patch are hidden together, removing their mutual context. MSE also treats bytes as numerical values and favors a conditional mean under uncertainty, although traffic bytes are discrete symbols. NetMamba similarly masks groups of bytes within a stride~\cite{wang2024netmamba}. These observations suggest that current contextual pretraining methods do not fully exploit local byte dependencies, motivating byte-level input and prediction objectives that preserve informative neighboring context.

\subsection{Byte-Level Contextual Models}
Building on this analysis, we develop two byte-level models designed to exploit neighboring-byte dependencies and evaluate whether better use of this local structure improves pretrained representations and downstream classification.

\textbf{ByteBERT.} We adapt the ET-BERT formulation so that each byte occupies one token position. Single-byte targets remove overlapping copies and allow neighboring unmasked bytes to serve as context.

For raw traffic $\mathcal{T}=(t_1,\ldots,t_{N_0})$, every byte occupies one token position, giving $\mathbf{u}=(u_1,\ldots,u_{N_0})$ with $u_i=t_i$. The encoder $f_w$ produces token representations and a pooled global representation
\begin{equation}
    \mathbf{R}_w=f_w(\mathbf{u})\in\mathbb{R}^{N_0\times d_w},\quad
    \overline{\mathbf{R}}_w=\mathrm{Pool}(\mathbf{R}_w)\in\mathbb{R}^{d_w}.
    \label{eq:rep_sb}
\end{equation}
For sample $i$, each valid position is independently selected with probability $p_{\mathrm{mask}}=0.15$ to form $\mathcal{M}^{(i)}$. The corrupted sequence is $\tilde{\mathbf{u}}^{(i)}$. For a batch of $B$ samples and $N_{\mathrm{mask}}=\sum_i|\mathcal{M}^{(i)}|$, the objective is
\begin{equation}
    \mathcal{L}_{\mathrm{MLM}}^w=-\frac{1}{N_{\mathrm{mask}}}
    \sum_{i=1}^{B}\sum_{k\in\mathcal{M}^{(i)}}
    \log P_{\theta_w}(u_k^{(i)}\mid\tilde{\mathbf{u}}^{(i)}).
    \label{eq:byte_mlm}
\end{equation}
This formulation exposes neighboring original byte positions to the objective whenever they remain uncorrupted. We match the BERT~\cite{devlin2019bert} backbone configuration and training schedule to ET-BERT, while only the vocabulary and prediction units differ.

\textbf{TrafficDCNet.} ByteBERT fixes the input unit at one byte. We further consider whether a model can learn how to group bytes as part of representation learning. Inspired by tokenizer-free hierarchical sequence models~\cite{hwang2025dynamic,slagle2024spacebyte,videau2025bytes}, TrafficDCNet embeds raw bytes and learns content-dependent boundaries through dynamic chunking. This design can adapt its internal resolution to the heterogeneous structure of visible fields and ciphertext without a fixed multi-byte vocabulary.

\begin{figure}[!tb]
    \centering
    \includegraphics[width=\linewidth]{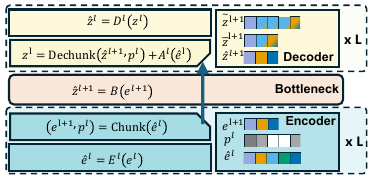}
    \caption{TrafficDCNet with one dynamic-chunking stage.}
    \label{fig:hnet}
\end{figure}

As shown in Fig.~\ref{fig:hnet}, its U-Net encoder~\cite{ronneberger2015u} progressively compresses the byte sequence. A causal Mamba-2 bottleneck~\cite{dao2024transformers} processes the compressed states, and a decoder restores the original byte resolution through dechunking and skip connections. Appendix~\ref{app:trafficdc} provides the full architecture, dynamic chunking and dechunking operations, and ratio regularizer.

TrafficDCNet is pretrained by next-byte prediction. With $t_{\leq i}=(t_1,\ldots,t_i)$, its objective is
\begin{equation}
    \begin{aligned}
        \mathcal{L}_{\mathrm{NBP}} & =-\frac{1}{N_0-1}\sum_{i=1}^{N_0-1}
        \log P_{\theta_b}(t_{i+1}\mid t_{\leq i}),                                                   \\
        \mathcal{L}_{\mathrm{DC}}  & =\mathcal{L}_{\mathrm{NBP}}+\alpha\mathcal{L}_{\mathrm{ratio}}.
    \end{aligned}
    \label{eq:dc_loss}
\end{equation}
The ratio loss regularizes dynamic chunking toward the target retained fractions. The encoder $f_b$ returns
\begin{equation}
    \mathbf{R}_b=f_b(\mathcal{T})\in\mathbb{R}^{N_0\times d_b},\quad
    \overline{\mathbf{R}}_b=\mathrm{Pool}(\mathbf{R}_b)\in\mathbb{R}^{d_b}.
    \label{eq:rep_dc}
\end{equation}

ByteBERT uses bidirectional masked prediction, while TrafficDCNet combines causal next-byte prediction with adaptive grouping. Both retain byte-resolution outputs, making them suitable for examining local context with different views.

\subsection{Classification and Representation Quality}
\begin{table*}[!t]
    \centering
    \caption{Flow-level classification with contextual pretraining on CSTNET and ISCX-Direct. Acc and Macro-F1 are percentages, reported as mean $\pm$ standard deviation.}
    \label{tab:context_perf}
    \setlength{\tabcolsep}{3pt}
    \renewcommand{\arraystretch}{1}
    \newcommand{\result}[2]{#1\ensuremath{\,\pm\,}#2}
    {\small\begin{tabular*}{\textwidth}{@{\extracolsep{\fill}}lcccccccc@{}}
            \toprule
            \multirow{3}{*}{Method} & \multicolumn{4}{c}{CSTNET} & \multicolumn{4}{c}{ISCX-Direct} \\
            \cmidrule(lr){2-5}\cmidrule(lr){6-9}
            & \multicolumn{2}{c}{Frozen} & \multicolumn{2}{c}{Unfrozen} & \multicolumn{2}{c}{Frozen} & \multicolumn{2}{c}{Unfrozen} \\
            \cmidrule(lr){2-3}\cmidrule(lr){4-5}\cmidrule(lr){6-7}\cmidrule(lr){8-9}
            & Acc & Macro-F1 & Acc & Macro-F1 & Acc & Macro-F1 & Acc & Macro-F1 \\
            \midrule
            ET-BERT & \result{17.89}{1.31} & \result{11.49}{1.81} & \result{62.56}{0.87} & \result{60.04}{0.68} & \result{23.20}{1.68} & \result{3.77}{1.58} & \result{71.00}{1.49} & \result{56.85}{1.55} \\
            YaTC & \result{38.04}{1.04} & \result{35.79}{1.74} & \result{74.52}{0.55} & \result{74.42}{1.27} & \result{65.68}{0.74} & \result{55.08}{1.06} & \result{77.22}{1.69} & \result{61.31}{2.00} \\
            NetMamba & \result{14.40}{1.38} & \result{11.80}{0.87} & \result{79.44}{0.99} & \result{77.10}{0.85} & \result{32.07}{0.57} & \result{24.45}{1.78} & \result{67.48}{0.74} & \result{65.57}{0.80} \\
            \midrule
            ByteBERT & \result{53.20}{0.90} & \result{48.91}{0.55} & \result{\textbf{88.62}}{1.27} & \result{\textbf{86.16}}{1.65} & \result{\textbf{71.92}}{1.81} & \result{62.55}{1.24} & \result{\textbf{86.49}}{1.52} & \result{\textbf{77.39}}{0.65} \\
            TrafficDCNet & \result{\textbf{62.10}}{1.14} & \result{\textbf{58.80}}{1.71} & \result{88.24}{0.68} & \result{85.98}{1.66} & \result{71.33}{1.40} & \result{\textbf{63.70}}{1.49} & \result{78.44}{1.08} & \result{67.74}{1.88} \\
            \bottomrule
        \end{tabular*}}
\end{table*}

\begin{figure}[!tb]
    \centering
    \includegraphics[width=\linewidth]{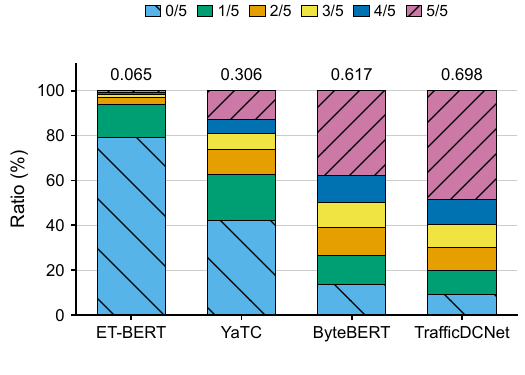}
    \caption{5NN-purity of frozen CSTNET flow representations under different modeling granularities. Each category gives the number of same-class neighbors among five nearest neighbors. Values above the bars denote mean purity.}
    \label{fig:5nn_granularity}
\end{figure}

\begin{figure}[!tb]
    \centering
    \subfloat[ET-BERT]{\includegraphics[width=0.48\linewidth]{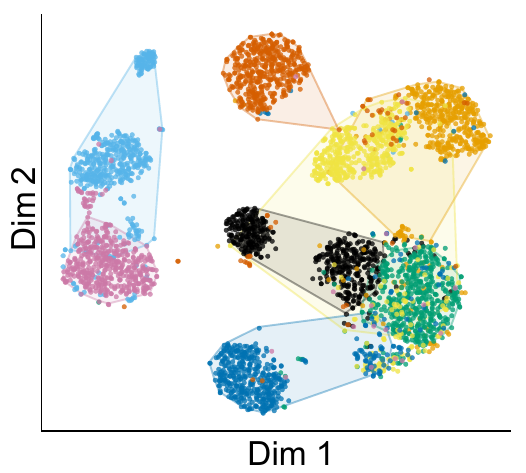}}
    \hfill
    \subfloat[YaTC]{\includegraphics[width=0.48\linewidth]{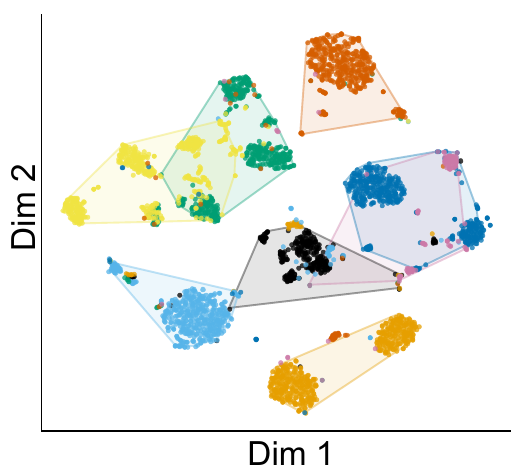}}\\
    \subfloat[ByteBERT]{\includegraphics[width=0.48\linewidth]{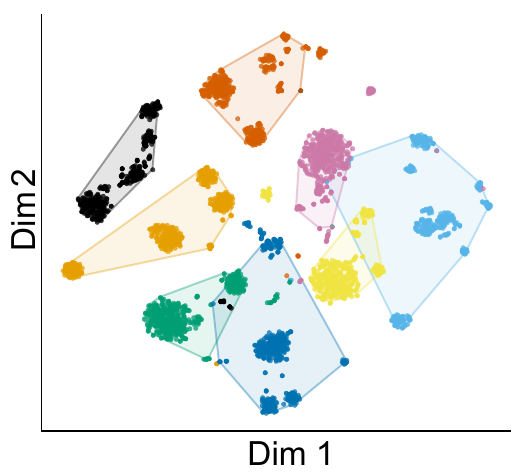}}
    \hfill
    \subfloat[TrafficDCNet]{\includegraphics[width=0.48\linewidth]{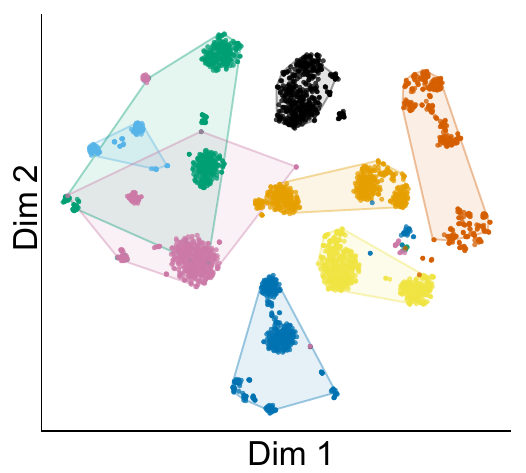}}
    \caption{t-SNE projections of frozen CSTNET flow representations. Each panel uses a model-specific subset of eight classes. Colors indicate class membership within each panel.}
    \label{fig:tsne_granularity}
\end{figure}

\begin{figure*}[!t]
    \centering
    \subfloat[YaTC]{\begin{minipage}[t][0.25\textwidth][t]{0.20\textwidth}\vspace{0pt}\centering
            \begin{tabular}[t]{@{}c@{\hspace{2pt}}c@{}}
                \includegraphics[width=0.46\linewidth]{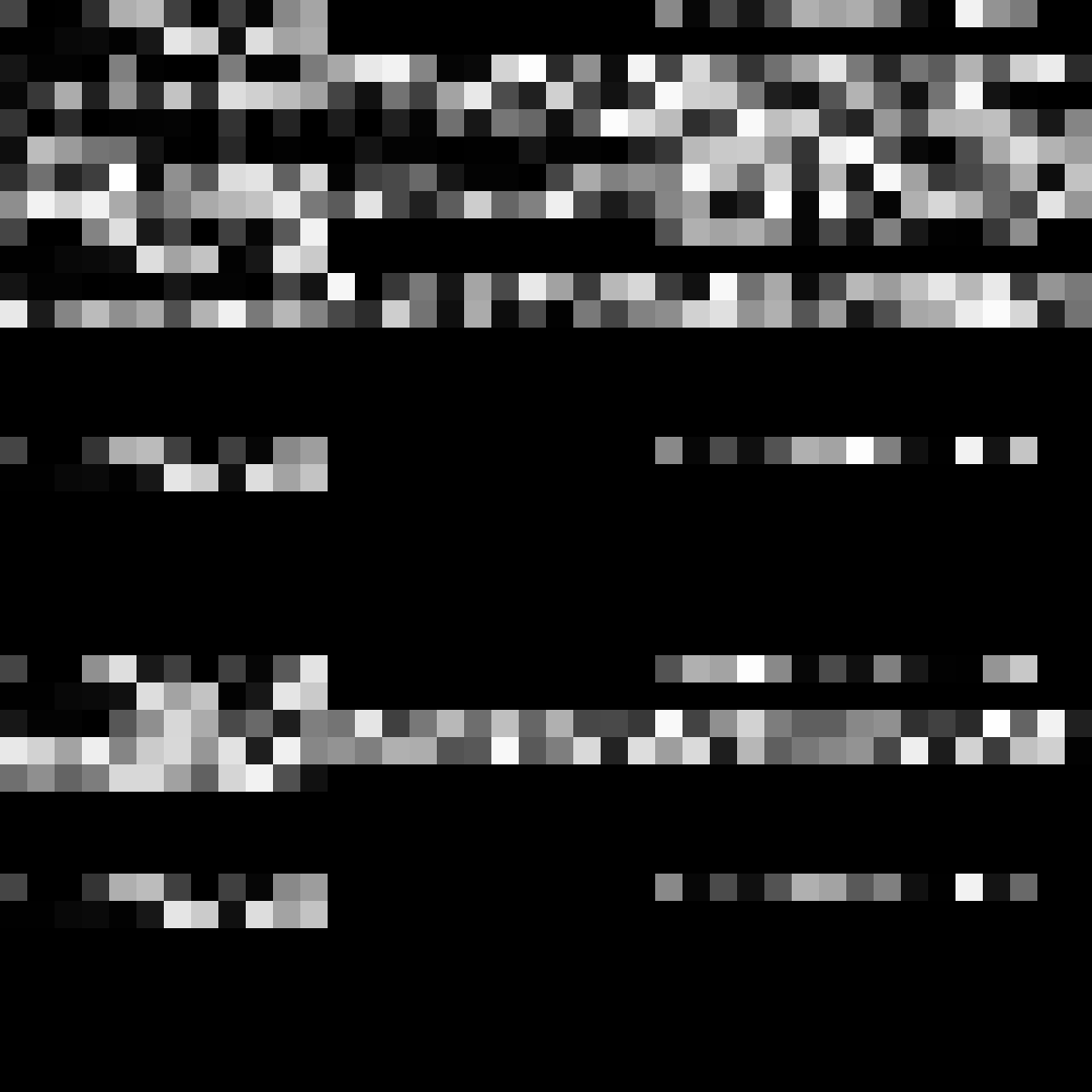}       &
                \includegraphics[width=0.46\linewidth]{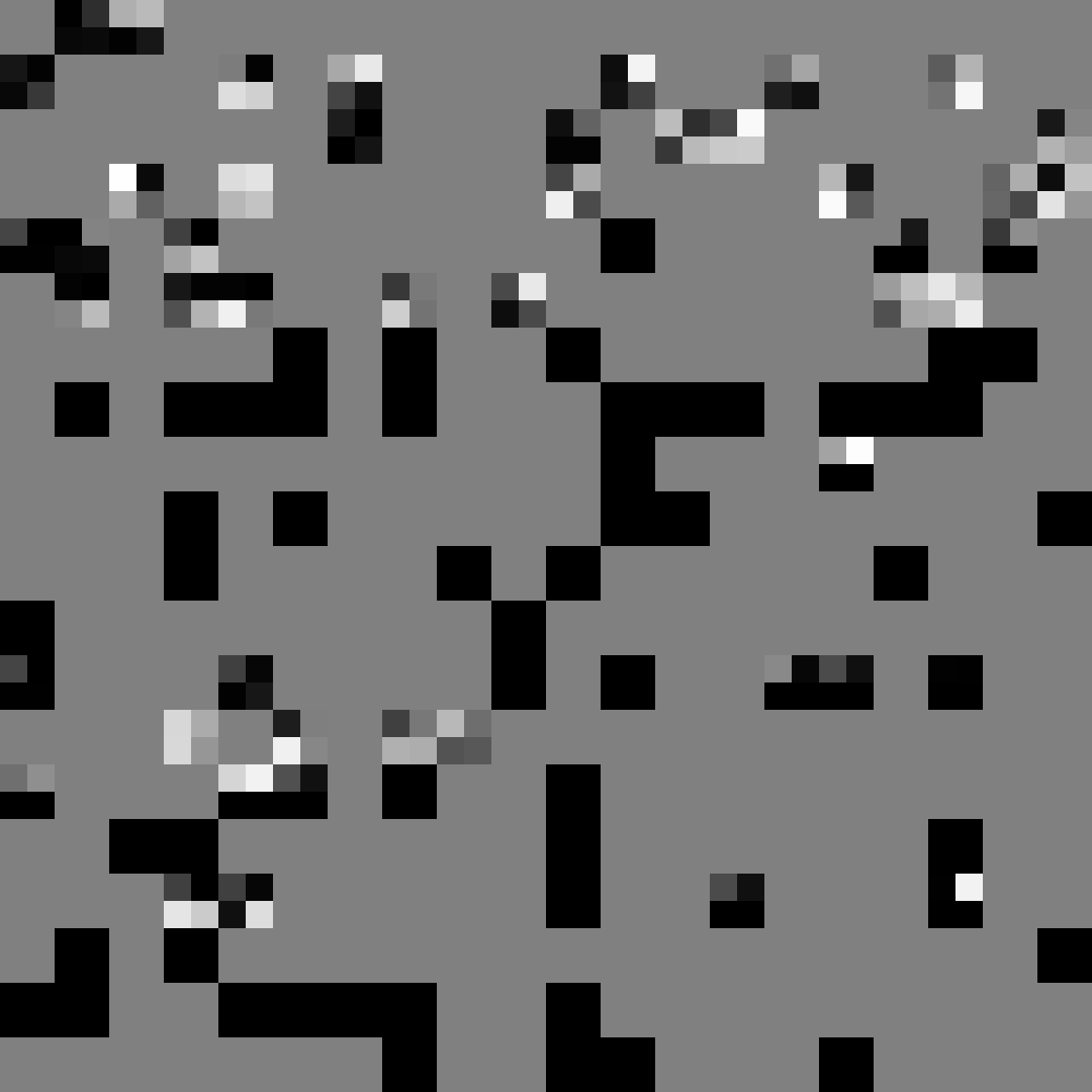}           \\
                \includegraphics[width=0.46\linewidth]{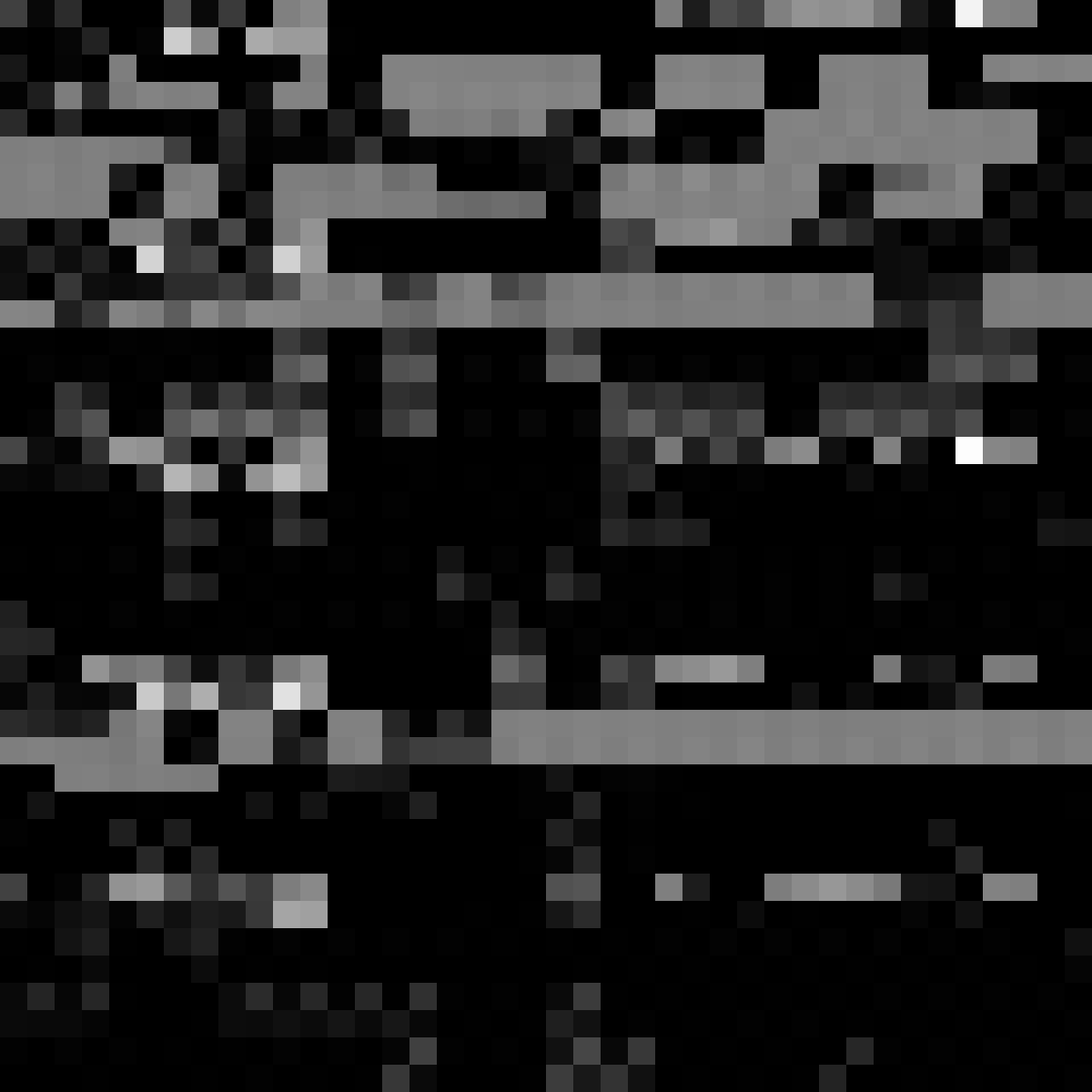} &
                \includegraphics[width=0.46\linewidth]{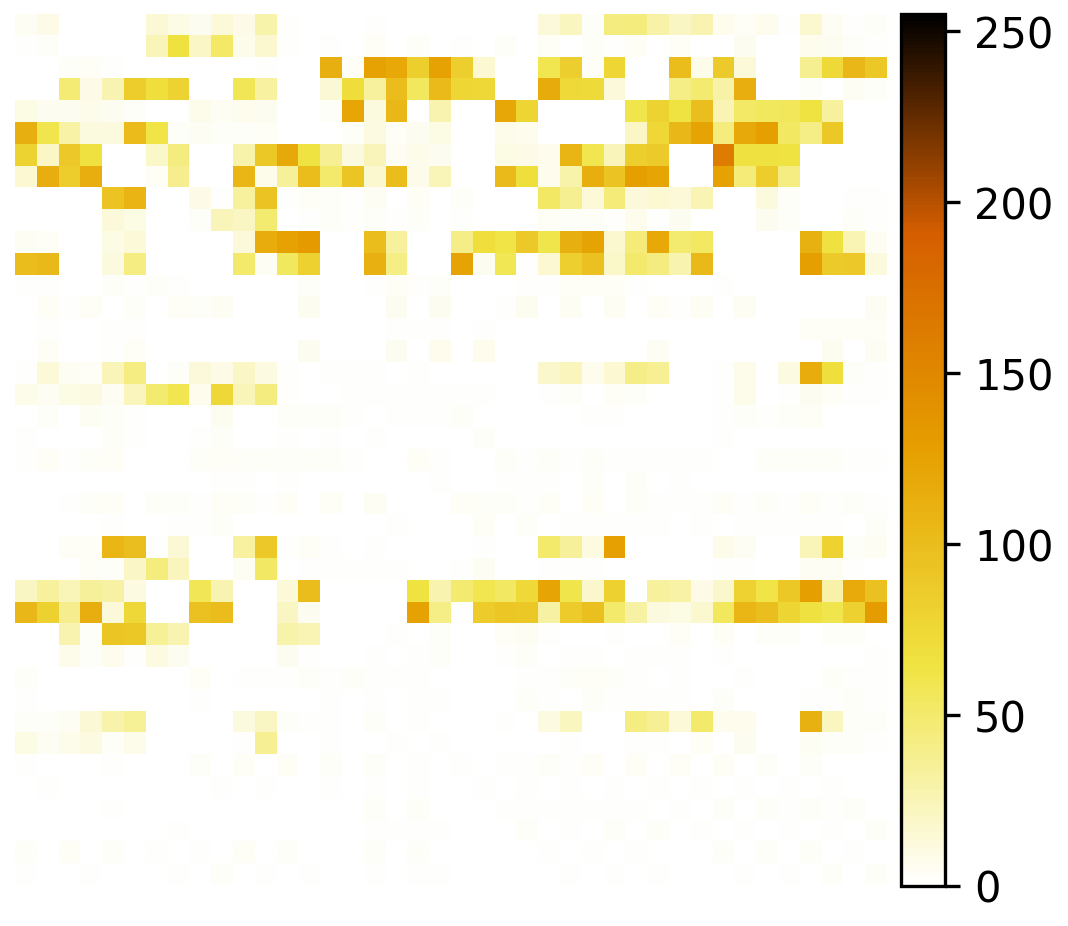}
            \end{tabular}\end{minipage}}
    \hfill
    \subfloat[ET-BERT]{\begin{minipage}[t][0.25\textwidth][t]{0.37\textwidth}\vspace{0pt}\centering
            \includegraphics[width=\linewidth]{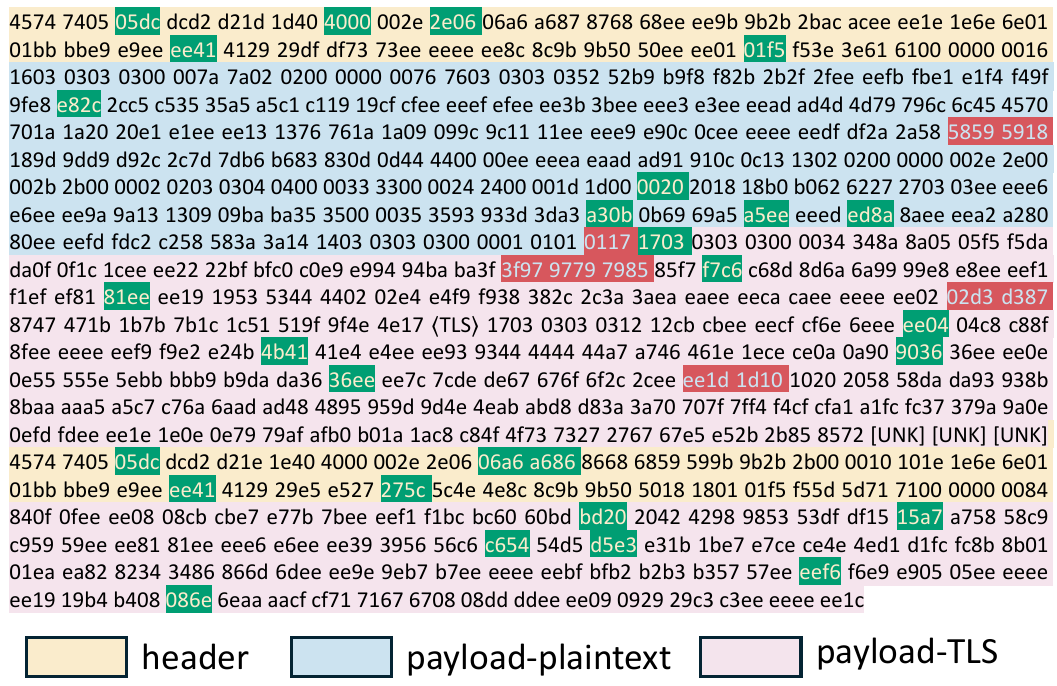}\end{minipage}}
    \hfill
    \subfloat[ByteBERT]{\begin{minipage}[t][0.25\textwidth][t]{0.37\textwidth}\vspace{0pt}\centering
            \includegraphics[width=\linewidth]{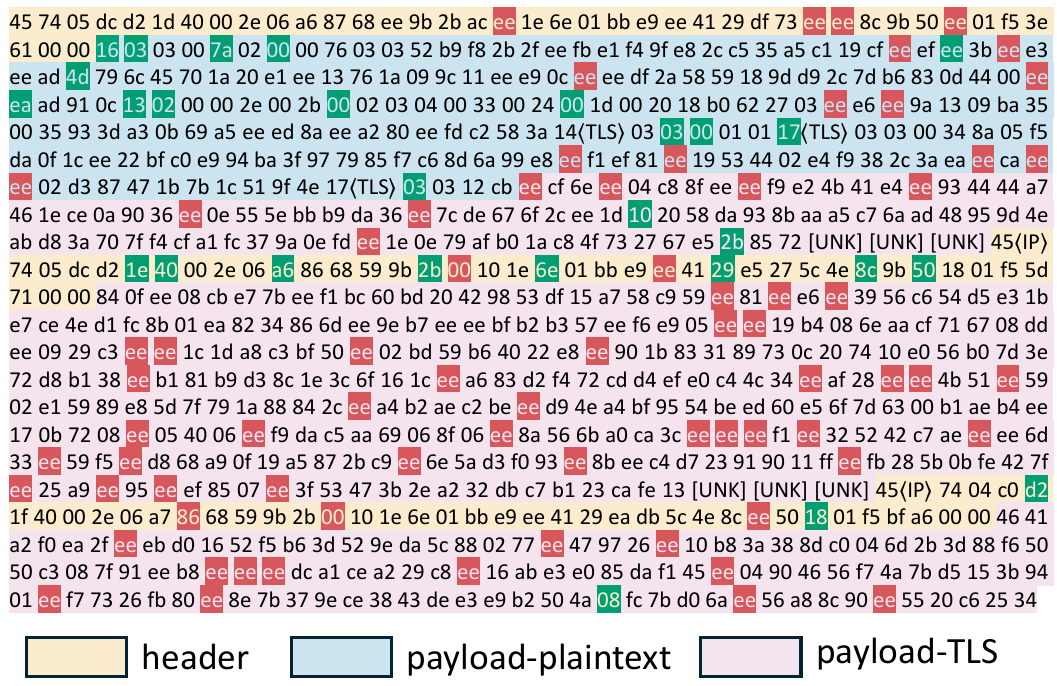}\end{minipage}}
    \caption{Qualitative pretraining examples. YaTC shows original, masked, reconstructed, and error images from top left to bottom right. In the byte examples, yellow marks headers, blue marks protocol-visible payload, and magenta marks TLS ciphertext. Green and red mark correct and incorrect predictions.}
    \label{fig:granularity_viz}
\end{figure*}

Table~\ref{tab:context_perf} compares the five contextual models on flow-level CSTNET and ISCX-Direct. ByteBERT and TrafficDCNet both improve frozen Macro-F1 over ET-BERT, YaTC, and NetMamba. On CSTNET, ByteBERT reaches 48.91\% and TrafficDCNet reaches 58.80\%, compared with 35.79\% for the strongest of these baselines. On ISCX-Direct, their scores are 62.55\% and 63.70\%, compared with 55.08\%. After fine-tuning, ByteBERT achieves 86.16\% Macro-F1 on CSTNET and 77.39\% on ISCX-Direct. These results support byte-level modeling as a way to use the available contextual signal.

We also measure the class consistency of frozen representations. A sample's 5NN-purity is the fraction of its five nearest neighbors that share its class label. Fig.~\ref{fig:5nn_granularity} shows mean purity increasing from 0.065 for ET-BERT and 0.306 for YaTC to 0.617 for ByteBERT and 0.698 for TrafficDCNet. This improvement indicates that the byte-level encoders organize samples into more class-consistent neighborhoods before task-specific encoder adaptation.
A similar pattern appears in the t-SNE projections~\cite{van2008visualizing} in Fig.~\ref{fig:tsne_granularity}. ByteBERT and TrafficDCNet exhibit compact local groups consistent with their higher purity and frozen classification scores.

We further visualize the models' predictions on their pretraining tasks in Fig.~\ref{fig:granularity_viz}. YaTC minimizes MSE and achieves a low error of 0.0777, yet its reconstructions largely collapse toward mean byte values and fail to reproduce the original patterns at masked positions. ET-BERT recovers many masked tokens, but its sliding-window tokenization leaves copies of the target bytes in neighboring tokens. This shortcut allows accurate token recovery with little need to learn dependencies between distinct original bytes, consistent with its weak frozen classification performance and low 5NN-purity. ByteBERT recovers part of the protocol-visible content through individual-byte prediction without overlapping target copies. Its higher frozen classification scores and 5NN-purity indicate that this contextual prediction produces more useful representations.

\section{Beyond Contextual Prediction through Cross-View Alignment}\label{sec:q3}
\subsection{Aligning Complementary Byte-Level Views}
Section~\ref{sec:q1} identifies a source of contextual supervision, and Section~\ref{sec:q2} shows how byte-level models can use it more effectively. Yet full-payload mutual information remains small because much of the observed sequence is encrypted. This raises the broader question of whether other pretraining paradigms can provide useful supervision when contextual prediction offers limited signal.
One alternative is contrastive learning, which derives supervision from correspondence between different views of the same instance rather than from contextual byte prediction~\cite{he2020momentum,li2021align}. Such views are often constructed through transformations that preserve the instance's semantic content. For encrypted traffic, however, these transformations are difficult to design because modifying bytes may disrupt protocol structure or remove application-relevant information, while encryption obscures the content needed to assess whether semantics are preserved. We therefore build on the two byte-level encoders developed in Section~\ref{sec:q2}, using their distinct modeling biases to construct complementary views of the same unmodified traffic sample. ByteBERT learns bidirectional dependencies through masked prediction over individual bytes, whereas TrafficDCNet combines causal next-byte prediction with adaptive byte grouping. These distinct modeling choices provide complementary views for alignment.

AlignTraffic aligns the global representations of these views and fuses their byte-level features. Appendix~\ref{app:align} provides the projection and fusion formulations and full definitions of the contrastive, matching, and MLM objectives.

\begin{figure}[!tb]
    \centering
    \includegraphics[width=\linewidth]{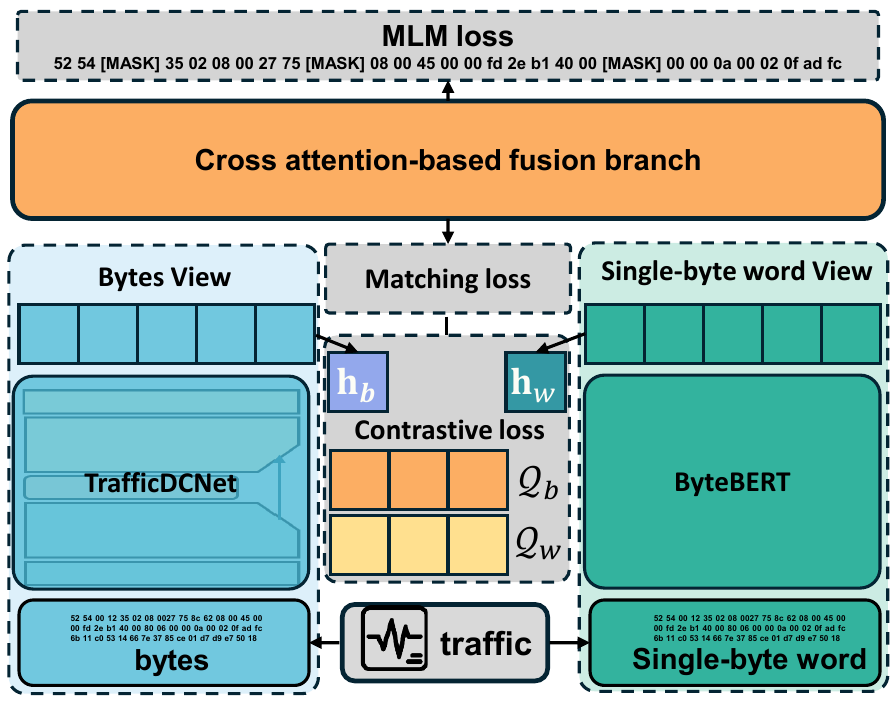}
    \caption{Architecture of AlignTraffic.}
    \label{fig:align}
\end{figure}

Projection heads map the pooled features to normalized embeddings $\mathbf{h}_w$ and $\mathbf{h}_b$ in a shared space. A contrastive objective $\mathcal{L}_{\mathrm{C}}$ brings together the two representations of the same sample relative to those of different samples, using momentum encoders and feature queues~\cite{he2020momentum,li2021align}. A fusion encoder applies cross-attention to the token-level representations, which share the original byte resolution. A matching objective $\mathcal{L}_{\mathrm{M}}$ trains the fused features to distinguish same-sample pairs from hard negative pairs.

Joint training also retains contextual supervision. The auxiliary fused-view MLM objective predicts bytes at synchronized corrupted positions in the two branches. With MLM weight $\eta$ and ratio-loss weight $\alpha$, the complete objective is
\begin{equation}
    \mathcal{L}_{\mathrm{Align}}=
    \mathcal{L}_{\mathrm{C}}+\mathcal{L}_{\mathrm{M}}
    +\eta\mathcal{L}_{\mathrm{MLM}}^{wb}
    +\alpha\mathcal{L}_{\mathrm{ratio}}.
    \label{eq:loss_align}
\end{equation}
We first pretrain ByteBERT and TrafficDCNet independently, then initialize AlignTraffic from those encoders and optimize the joint objective. The correspondence objectives organize the complementary features around sample identity, while fused-view MLM preserves their connection to the observed bytes.

\subsection{Classification with Cross-View Alignment}
We next examine whether the additional supervision from cross-view alignment improves classification beyond contextual pretraining alone and whether AlignTraffic can outperform shallow learning, whose effectiveness was emphasized by Zhao et al.~\cite{10.1145/3718958.3750498}. Tables~\ref{tab:flow_perf} and~\ref{tab:packet_perf} compare AlignTraffic with shallow learning, Pcap-Encoder, and its two constituent encoders. The shallow baseline selects among Random Forest, XGBoost, and LightGBM using explicit TCP/IP and TLS features, including record statistics and visible ServerHello fields. Appendix~\ref{app:shallow} details the feature groups. For Pcap-Encoder, we report only unfrozen flow-level results. Following the original evaluation procedure, we obtain flow labels by majority voting over packet predictions~\cite{10.1145/3718958.3750498}.

\begin{table*}[!t]
    \centering
    \caption{Flow-level classification on CSTNET and ISCX-Direct. Acc and Macro-F1 are percentages, reported as mean $\pm$ standard deviation.}
    \label{tab:flow_perf}
    \setlength{\tabcolsep}{3pt}
    \renewcommand{\arraystretch}{1}
    \newcommand{\result}[2]{#1\ensuremath{\,\pm\,}#2}
    {\small\begin{tabular*}{\textwidth}{@{\extracolsep{\fill}}lcccccccc@{}}
            \toprule
            \multirow{3}{*}{Method} & \multicolumn{4}{c}{CSTNET} & \multicolumn{4}{c}{ISCX-Direct} \\
            \cmidrule(lr){2-5}\cmidrule(lr){6-9}
            & \multicolumn{2}{c}{Frozen} & \multicolumn{2}{c}{Unfrozen} & \multicolumn{2}{c}{Frozen} & \multicolumn{2}{c}{Unfrozen} \\
            \cmidrule(lr){2-3}\cmidrule(lr){4-5}\cmidrule(lr){6-7}\cmidrule(lr){8-9}
            & Acc & Macro-F1 & Acc & Macro-F1 & Acc & Macro-F1 & Acc & Macro-F1 \\
            \midrule
            Shallow learning & -- & -- & \result{78.11}{1.66} & \result{58.50}{1.50} & -- & -- & \result{86.09}{0.85} & \result{69.08}{1.38} \\
            Pcap-Encoder & -- & -- & \result{71.35}{1.48} & \result{61.88}{1.21} & -- & -- & \result{84.75}{0.86} & \result{68.52}{1.01} \\
            \midrule
            ByteBERT & \result{53.20}{0.90} & \result{48.91}{0.55} & \result{88.62}{1.27} & \result{86.16}{1.65} & \result{71.92}{1.81} & \result{62.55}{1.24} & \result{86.49}{1.52} & \result{77.39}{0.65} \\
            TrafficDCNet & \result{62.10}{1.14} & \result{58.80}{1.71} & \result{88.24}{0.68} & \result{85.98}{1.66} & \result{71.33}{1.40} & \result{63.70}{1.49} & \result{78.44}{1.08} & \result{67.74}{1.88} \\
            AlignTraffic & \result{\textbf{67.08}}{1.23} & \result{\textbf{64.22}}{0.95} & \result{\textbf{90.35}}{0.70} & \result{\textbf{89.20}}{1.88} & \result{\textbf{76.34}}{0.68} & \result{\textbf{65.04}}{0.97} & \result{\textbf{90.36}}{0.93} & \result{\textbf{83.92}}{1.01} \\
            \bottomrule
        \end{tabular*}}
\end{table*}

\begin{table*}[!t]
    \centering
    \caption{Packet-level classification on ISCX-Direct and ISCX-VPN. Acc and Macro-F1 are percentages, reported as mean $\pm$ standard deviation.}
    \label{tab:packet_perf}
    \setlength{\tabcolsep}{3pt}
    \renewcommand{\arraystretch}{1}
    \newcommand{\result}[2]{#1\ensuremath{\,\pm\,}#2}
    {\small\begin{tabular*}{\textwidth}{@{\extracolsep{\fill}}lcccccccc@{}}
            \toprule
            \multirow{3}{*}{Method} & \multicolumn{4}{c}{ISCX-Direct} & \multicolumn{4}{c}{ISCX-VPN} \\
            \cmidrule(lr){2-5}\cmidrule(lr){6-9}
            & \multicolumn{2}{c}{Frozen} & \multicolumn{2}{c}{Unfrozen} & \multicolumn{2}{c}{Frozen} & \multicolumn{2}{c}{Unfrozen} \\
            \cmidrule(lr){2-3}\cmidrule(lr){4-5}\cmidrule(lr){6-7}\cmidrule(lr){8-9}
            & Acc & Macro-F1 & Acc & Macro-F1 & Acc & Macro-F1 & Acc & Macro-F1 \\
            \midrule
            Shallow learning & -- & -- & \result{93.07}{1.12} & \result{68.37}{1.44} & -- & -- & \result{94.43}{0.78} & \result{59.35}{1.31} \\
            Pcap-Encoder & \result{96.97}{0.68} & \result{39.94}{1.39} & \result{98.66}{0.57} & \result{62.67}{1.17} & \result{\textbf{97.01}}{0.63} & \result{65.01}{1.08} & \result{97.45}{0.55} & \result{66.56}{1.14} \\
            \midrule
            ByteBERT & \result{96.99}{0.61} & \result{56.66}{1.57} & \result{97.80}{0.66} & \result{66.43}{1.20} & \result{89.76}{0.92} & \result{60.17}{1.34} & \result{95.71}{0.70} & \result{75.92}{1.03} \\
            TrafficDCNet & \result{\textbf{97.40}}{0.56} & \result{64.83}{1.19} & \result{99.33}{0.51} & \result{71.18}{1.08} & \result{93.95}{0.81} & \result{72.70}{1.27} & \result{96.88}{0.60} & \result{76.21}{0.98} \\
            AlignTraffic & \result{\textbf{97.40}}{0.54} & \result{\textbf{64.87}}{1.02} & \result{\textbf{99.44}}{0.50} & \result{\textbf{74.34}}{0.89} & \result{95.91}{0.73} & \result{\textbf{74.31}}{1.10} & \result{\textbf{97.73}}{0.58} & \result{\textbf{83.36}}{0.84} \\
            \bottomrule
        \end{tabular*}}
\end{table*}

After fine-tuning, AlignTraffic achieves the highest Macro-F1 among the compared methods on all four public tasks. On flow-level CSTNET and ISCX-Direct, it reaches 89.20\% and 83.92\%, exceeding shallow learning by 30.70 and 14.84 percentage points, respectively. It also outperforms Pcap-Encoder on both flow-level tasks. On packet-level ISCX-Direct and ISCX-VPN, its gains over shallow learning are 5.97 and 24.01 percentage points.

Among the reported frozen results, AlignTraffic achieves the highest Macro-F1 on all four public tasks, outperforming both constituent encoders. On CSTNET, alignment raises the score from 48.91\% for ByteBERT and 58.80\% for TrafficDCNet to 64.22\%. Packet-level ISCX-VPN shows the same progression, from 60.17\% and 72.70\% to 74.31\%. The improvement with a fixed downstream encoder demonstrates that alignment strengthens the pretrained representation.

\subsection{Representation Structure and Optimization}
\begin{figure}[!tb]
    \centering
    \subfloat[5NN-purity]{\includegraphics[width=0.60\linewidth]{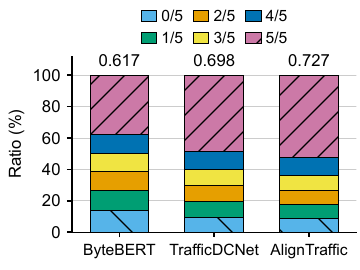}\label{fig:5nn_complementarity}}
    \hfill
    \subfloat[AlignTraffic t-SNE]{\raisebox{0.03377\linewidth}[0.45\linewidth][0pt]{\includegraphics[width=0.37\linewidth]{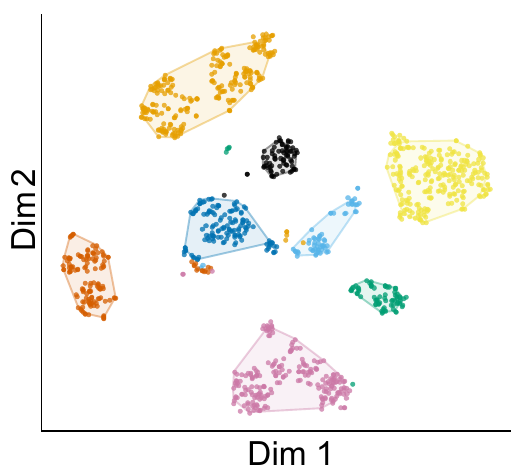}}\label{fig:tsne_complementarity}}
    \caption{Frozen CSTNET flow representations after cross-view alignment. Panel (a) compares ByteBERT, TrafficDCNet, and AlignTraffic using the purity categories in Fig.~\ref{fig:5nn_granularity}. Panel (b) projects eight classes using t-SNE.}
    \label{fig:alignment_representation}
\end{figure}

Fig.~\ref{fig:5nn_complementarity} compares the two byte-level encoders with AlignTraffic. Mean 5NN-purity increases from 0.617 for ByteBERT and 0.698 for TrafficDCNet to 0.727 after alignment. The aligned representation has the highest mean purity among the five encoders in the neighborhood analysis. Fig.~\ref{fig:tsne_complementarity} illustrates the resulting class structure, with separated groups in the projected features.

\begin{figure*}[!t]
    \centering
    \begin{minipage}[t]{0.49\textwidth}
        \vspace{0pt}\centering
        \includegraphics[width=\linewidth]{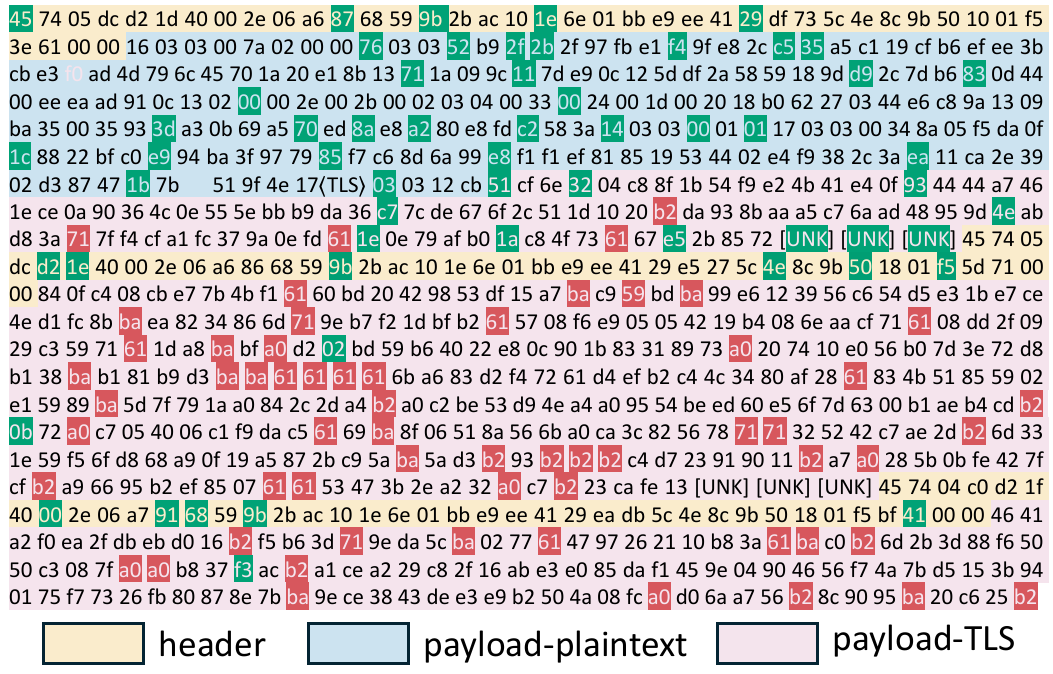}
    \end{minipage}\hfill
    \begin{minipage}[t]{0.49\textwidth}
        \vspace{0pt}\centering
        \includegraphics[width=\linewidth]{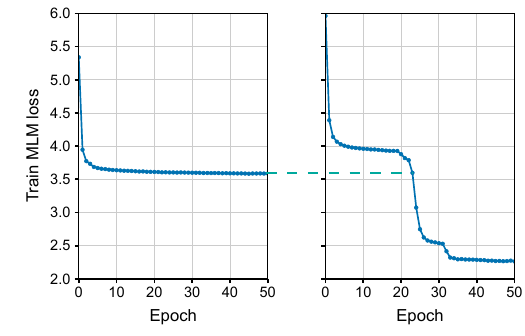}
    \end{minipage}\\[-1pt]
    \subfloat[AlignTraffic MLM predictions]{\makebox[0.49\textwidth]{\rule{0pt}{0pt}}}
    \hfill
    \subfloat[ByteBERT MLM loss]{\makebox[0.235\textwidth]{\rule{0pt}{0pt}}}
    \hfill
    \subfloat[AlignTraffic MLM loss]{\makebox[0.235\textwidth]{\rule{0pt}{0pt}}}
    \caption{Masked-byte prediction and MLM training. In (a), yellow, blue, and magenta indicate headers, protocol-visible payload, and TLS ciphertext. Green and red indicate correct and incorrect predictions. Panels (b) and (c) count epochs within their respective pretraining stages. The continuous green dashed line connects the reference loss across the two panels.}
    \label{fig:loss}
\end{figure*}

Fig.~\ref{fig:loss}(a) shows AlignTraffic recovering protocol-visible byte structure. The two loss curves in Fig.~\ref{fig:loss}(b) and (c) connect independent ByteBERT pretraining to alignment training. ByteBERT's MLM loss decreases early and then stabilizes, whereas fused-view MLM continues to improve during AlignTraffic training. The continuous reference line makes this progression visible across the two stages. Summarising the frozen classification and purity results, these observations show that cross-view training improves how the encoders use the available information.

\subsection{Objective Ablation}
\begin{table*}[!t]
    \centering
    \caption{Flow-level objective ablation on CSTNET and ISCX-Direct. Acc and Macro-F1 are percentages.}
    \label{tab:align_ablation}
    \setlength{\tabcolsep}{3pt}
    \renewcommand{\arraystretch}{1}
    \newcommand{\result}[2]{#1\ensuremath{\,\pm\,}#2}
    {\small\begin{tabular*}{\textwidth}{@{\extracolsep{\fill}}lcccccccc@{}}
            \toprule
            \multirow{3}{*}{Method} & \multicolumn{4}{c}{CSTNET} & \multicolumn{4}{c}{ISCX-Direct} \\
            \cmidrule(lr){2-5}\cmidrule(lr){6-9}
            & \multicolumn{2}{c}{Frozen} & \multicolumn{2}{c}{Unfrozen} & \multicolumn{2}{c}{Frozen} & \multicolumn{2}{c}{Unfrozen} \\
            \cmidrule(lr){2-3}\cmidrule(lr){4-5}\cmidrule(lr){6-7}\cmidrule(lr){8-9}
            & Acc & Macro-F1 & Acc & Macro-F1 & Acc & Macro-F1 & Acc & Macro-F1 \\
            \midrule
            AlignTraffic & \textbf{67.08} & \textbf{64.22} & \textbf{90.35} & \textbf{89.20} & \textbf{76.34} & \textbf{65.04} & \textbf{90.36} & \textbf{83.92} \\
            w/o MLM & 35.84 & 29.97 & 83.03 & 80.76 & 41.59 & 26.44 & 68.60 & 55.50 \\
            w/o contrastive & 58.75 & 53.87 & 82.55 & 79.07 & 72.16 & 62.58 & 79.86 & 71.03 \\
            w/o matching & 55.84 & 49.97 & 83.03 & 80.76 & 70.62 & 58.42 & 77.37 & 65.77 \\
            w/o alignment & 51.11 & 46.89 & 84.23 & 81.94 & 57.70 & 48.44 & 80.09 & 65.58 \\
            \bottomrule
        \end{tabular*}}
\end{table*}

We retain the two-branch framework and remove individual objectives to assess their contributions. Table~\ref{tab:align_ablation} shows that removing both alignment objectives reduces frozen Macro-F1 by 17.33 percentage points on CSTNET and 16.60 percentage points on ISCX-Direct. Removing either contrastive or matching supervision also lowers performance. MLM makes a complementary contribution, with its removal producing the largest reduction in frozen Macro-F1. The full objective therefore benefits from both byte-level contextual learning and cross-view correspondence.

\subsection{Industrial Classification and Geographic Transfer}
We further evaluate AlignTraffic on a private industrial flow-classification dataset collected from March 2025 to March 2026. It covers 165 applications across multiple devices and network conditions. Inputs retain payloads from flows with at least eight packets and remove packet headers. The training set contains approximately 1.38 million samples, and the test set contains approximately 2.83 million samples with an imbalanced deployment distribution. Appendix~\ref{app:industrial} describes the collection.

\begin{table}[!tb]
    \centering
    \small
    \setlength{\tabcolsep}{4pt}
    \renewcommand{\arraystretch}{0.95}
    \caption{Unfrozen flow-level classification on the industrial dataset. Acc and Macro-F1 are percentages.}
    \label{tab:private_results}
    \begin{tabular}{@{}lcc@{}}
        \toprule
        Method              & Acc            & Macro-F1       \\
        \midrule
        ET-BERT             & 67.87          & 48.21          \\
        YaTC                & 82.90          & 42.87          \\
        Industrial baseline & 92.15          & 42.38          \\
        AlignTraffic        & \textbf{94.54} & \textbf{51.26} \\
        \bottomrule
    \end{tabular}
\end{table}

Table~\ref{tab:private_results} shows that AlignTraffic reaches 94.54\% Acc and 51.26\% Macro-F1. It improves Macro-F1 over the industrial baseline by 8.88 percentage points and exceeds the compared public models. This extends the classification evidence to a larger application set and an industrial collection environment.

In a separate geographic transfer experiment, we train on 887,781 samples from a source region and evaluate on held-out traffic from Beijing and Xi'an without target-region retraining. The two test sets contain 165,144 and 24,626 samples, respectively.

\begin{table}[!tb]
    \centering
    \caption{Flow-level classification on traffic from Beijing and Xi'an without target-region retraining. Acc and Macro-F1 are percentages.}
    \label{tab:geographic}
    \small
    \setlength{\tabcolsep}{3pt}
    \begin{tabular*}{\columnwidth}{@{\extracolsep{\fill}}llrr@{}}
        \toprule
        Method & Region & Acc & Macro-F1\\
        \midrule
        Industrial baseline & Beijing & 94.79 & 95.42\\
        AlignTraffic & Beijing & \textbf{95.83} & \textbf{96.08}\\
        Improvement & Beijing & +1.04 & +0.66\\
        \midrule
        Industrial baseline & Xi'an & 81.41 & 83.95\\
        AlignTraffic & Xi'an & \textbf{87.51} & \textbf{88.69}\\
        Improvement & Xi'an & +6.10 & +4.74\\
        \bottomrule
    \end{tabular*}
\end{table}

Table~\ref{tab:geographic} shows that AlignTraffic improves Macro-F1 over the industrial baseline by 0.66 percentage points in Beijing and 4.74 percentage points in Xi'an.

\subsection{Computational Cost}
\begin{table}[!tb]
    \centering
    \caption{Computational cost in the offline evaluation. Parameters are in millions, latency in milliseconds, and memory in MB.}
    \label{tab:efficiency}
    \setlength{\tabcolsep}{3pt}
    {\small\begin{tabular*}{\columnwidth}{@{\extracolsep{\fill}}llrrr@{}}
            \toprule
            Method & Stage & \shortstack{Params\\(M)} & \shortstack{Latency\\(ms)} & \shortstack{Memory\\(MB)}\\
            \midrule
            \multirow{2}{*}{ET-BERT}&Inference&\multirow{2}{*}{160.67}&17.54&1013.15\\
            &Pretraining&&62.69&3122.84\\
            \multirow{2}{*}{Pcap-Encoder}&Inference&\multirow{2}{*}{222.90}&36.04&669.00\\
            &Pretraining&&94.69&4311.69\\
            \multirow{2}{*}{TrafficDCNet}&Inference&\multirow{2}{*}{249.32}&48.38&980.42\\
            &Pretraining&&199.67&2431.41\\
            \multirow{2}{*}{AlignTraffic}&Inference&\multirow{2}{*}{396.35}&62.82&3789.12\\
            &Pretraining&&380.07&9648.35\\
            \bottomrule
        \end{tabular*}}
\end{table}

Although the additional contrastive supervision improves AlignTraffic's representations and classification performance, its competitiveness in practice also depends on computational efficiency. Encoding each sample through two branches and fusing their features introduces additional computation and memory costs. Table~\ref{tab:efficiency} reports the parameter counts, latency, and memory usage measured in the offline evaluation. AlignTraffic uses 396.35 million parameters and takes 62.82 ms for inference, compared with 160.67 million parameters and 17.54 ms for ET-BERT. Its pretraining memory usage reaches 9648.35 MB. For deployments with tight latency or memory budgets, these costs must be weighed against the classification gains~\cite{jafri2024leo,parvez2025splidt}.

\section{Limits of Encrypted-Traffic Pretraining}\label{sec:q4}
\subsection{Application Information Concealed by Tor}
The preceding sections identify learnable structure in protocol-visible fields and demonstrate how byte-level modeling and cross-view alignment improve traffic representations. We now extend the evaluation to more restricted settings, like ISCX-Tor~\cite{icissp17}. Tor carries application streams through encrypted relay cells over TLS channels, concealing the inner application's protocol structure from an observer of the outer connection~\cite{torSpecifications}. This provides a setting with more restricted application-specific observations than direct TLS traffic.

We apply the same protocol-based filtering used for the ISCX datasets. The retained subset contains 1,846,917 packets, compared with 1,854,409 before filtering, and its payload entropy increases from 4.43 to 7.99 bits. With 68 independent retained flows, we evaluate classification at packet level using flow-disjoint partitions and the same identifier masking.

\begin{table}[!tb]
    \centering
    \caption{Packet-level classification on ISCX-Tor. Acc and Macro-F1 are percentages, reported as mean $\pm$ standard deviation.}
    \label{tab:tor_perf}
    \setlength{\tabcolsep}{3pt}
    \renewcommand{\arraystretch}{1}
    \newcommand{\result}[2]{\shortstack[c]{#1\\[-0.2ex]\ensuremath{\pm}\,#2}}
    {\small\begin{tabular*}{\columnwidth}{@{\extracolsep{\fill}}lcccc@{}}
            \toprule
            \multirow{3}{*}{Method} & \multicolumn{4}{c}{ISCX-Tor} \\
            \cmidrule(lr){2-5}
            & \multicolumn{2}{c}{Frozen} & \multicolumn{2}{c}{Unfrozen} \\
            \cmidrule(lr){2-3}\cmidrule(lr){4-5}
            & Acc & Macro-F1 & Acc & Macro-F1 \\
            \midrule
            ET-BERT & \result{53.42}{1.29} & \result{9.40}{1.84} & \result{55.14}{1.58} & \result{19.34}{1.11} \\
            YaTC & \result{3.99}{1.52} & \result{4.34}{1.79} & \result{11.00}{1.64} & \result{6.35}{1.36} \\
            NetMamba & \result{2.79}{1.88} & \result{2.95}{1.56} & \result{4.50}{1.73} & \result{1.58}{1.42} \\
            Pcap-Encoder & \result{\textbf{60.06}}{1.35} & \result{15.00}{1.62} & \result{\textbf{63.50}}{1.06} & \result{22.47}{1.49} \\
            Shallow learning & -- & -- & \result{8.86}{1.69} & \result{3.22}{1.91} \\
            \midrule
            ByteBERT & \result{59.65}{1.24} & \result{9.34}{1.71} & \result{62.60}{1.09} & \result{25.65}{1.37} \\
            TrafficDCNet & \result{21.44}{1.41} & \result{15.62}{1.55} & \result{44.66}{1.28} & \result{20.64}{1.63} \\
            AlignTraffic & \result{37.61}{1.33} & \result{\textbf{23.75}}{1.22} & \result{61.25}{1.04} & \result{\textbf{25.78}}{1.16} \\
            \bottomrule
        \end{tabular*}}
\end{table}

Table~\ref{tab:tor_perf} shows low Macro-F1 across the evaluated methods. AlignTraffic obtains the highest frozen and unfrozen scores, 23.75\% and 25.78\%, respectively, but reliable recognition across classes remains unresolved in this setting. The result identifies a boundary for the current byte-level representations when application structure is concealed by additional encrypted encapsulation.

\subsection{Information Available to Individual Packets}
Packet-level CSTNET restricts the observation in a different way. The flow-level interventions in Q1 show that classification-relevant structure is concentrated in the first retained packet. Individual later packets receive the same flow label while containing primarily encrypted payload.

\begin{table}[!tb]
    \centering
    \caption{Unfrozen packet-level classification on CSTNET. Acc and Macro-F1 are percentages.}
    \label{tab:cstnet-packet}
    {\small\begin{tabular}{lcc}
            \toprule
            \multirow{2}{*}{\textbf{Method}} & \multicolumn{2}{c}{CSTNET}                     \\
            \cmidrule(lr){2-3}
                                             & \textbf{Acc}               & \textbf{Macro-F1} \\
            \midrule
            ET-BERT                          & 25.41                      & 16.15             \\
            YaTC                             & 59.38                      & 33.28             \\
            NetMamba                         & 57.21                      & 29.33             \\
            Shallow learning                 & \textbf{71.06}             & \textbf{63.73}    \\
            \midrule
            ByteBERT                         & 60.71                      & 37.21             \\
            TrafficDCNet                     & 31.24                      & 23.31             \\
            AlignTraffic                     & 35.47                      & 24.97             \\
            \bottomrule
        \end{tabular}}
\end{table}

Table~\ref{tab:cstnet-packet} reports 37.21\% Macro-F1 for ByteBERT, 23.31\% for TrafficDCNet, and 24.97\% for AlignTraffic. Shallow learning reaches 63.73\%, showing the utility of explicit protocol features for this packet-level task. The comparison highlights the role of the observation unit. A representation that benefits from a multi-packet prefix may have substantially less class information available in an isolated packet.

\section{Related Work}\label{sec:related}
\subsection{Traffic Classification and Pretraining}
Traffic classification evolved from port matching and DPI to statistical features and raw-byte neural models as encryption reduced access to application content~\cite{haffner2005acas,karagiannis2005blinc,bernaille2007early,wang2017end,lotfollahi2020deep}. Self-supervised pretraining extends this progression by learning representations from unlabeled traces. PERT, ET-BERT, YaTC, NetMamba, and PacRep explore different input units and prediction objectives~\cite{he2020pert,10.1145/3485447.3512217,zhao2023yet,wang2024netmamba,meng2022packet}. Graph-based models and contrastive learning extend these approaches~\cite{zhang2023tfe,zhang2024one,lin2024cetp}. We relate model granularity and supervision to the byte dependencies that remain observable in encrypted traces.

\subsection{Evaluation Bias and Generalization}
Zhao et al.~\cite{10.1145/3718958.3750498} examine dataset preparation, flow-disjoint splitting, frozen evaluation, and protocol-aware pretraining. Their analysis motivates separating downstream classification accuracy from the quality of a pretrained representation. Related studies investigate outdated datasets and collection-dependent shortcuts~\cite{wickramasinghe2025sok,wang2026bias}. Rosetta shows that TCP dynamics can change the observed shape of TLS flows across environments, while data-drift studies document corresponding classifier sensitivity~\cite{xie2023rosetta,malekghaini2022data}. We use flow-disjoint partitions, identifier masking, and both frozen and unfrozen evaluations, and extend the comparison to industrial geographic transfer.

\subsection{Protocol-Visible Information}
Header-based classification and Pcap-Encoder demonstrate the value of observable protocol information~\cite{Cui_2022,yu2024novel,10.1145/3718958.3750498}. Akbari et al.~\cite{akbari2022traffic} show that encrypted-traffic classifiers can rely heavily on visible TLS handshake fields, motivating protocol-aware input design. \mbox{ET-BERT} also discusses ciphertext randomness as a challenge for traffic modeling~\cite{10.1145/3485447.3512217}. Our analysis distinguishes visible fields from ciphertext within retained payloads and connects their byte dependencies to classification through packet-specific interventions. ByteBERT and TrafficDCNet examine ways to exploit local structure, drawing on byte-level and tokenizer-free sequence modeling~\cite{slagle2024spacebyte,hwang2025dynamic,videau2025bytes}. AlignTraffic further uses their complementary views to provide supervision through sample correspondence.

\section{Conclusion}\label{sec:conclusion}
Encrypted traffic retains learnable structure in protocol-visible regions. Our entropy and mutual-information analysis identifies local byte dependencies, and CSTNET payload interventions connect the retained structure to classification. ByteBERT and TrafficDCNet exploit this signal through individual-byte prediction and adaptive grouping. AlignTraffic strengthens their representations with cross-view contrastive and matching supervision, improving public and industrial classification and supporting transfer across geographic collection settings. Evaluations with concealed application structure and individual-packet observations characterize the remaining boundaries. These findings establish a basis for encrypted-traffic pretraining that matches model design and supervision to the information available in observed traces.

\appendices
\section{TrafficDCNet Architecture}\label{app:trafficdc}
\subsection{Hierarchical encoder and decoder}
TrafficDCNet embeds the raw byte sequence $\mathcal{T}$ into $\mathbf{e}^0\in\mathbb{R}^{N_0\times d_0}$. At level $l$, $N_l$ is the sequence length and $d_l$ is the feature width. For $l=0,\ldots,L-1$, contextual transformation precedes dynamic chunking
\begin{equation}
    \begin{aligned}
        \mathbf{e}^0&=\mathrm{Emb}(\mathcal{T}),\\
        \hat{\mathbf{e}}^l&=E^l(\mathbf{e}^l),\\
        (\mathbf{e}^{l+1},\mathbf{p}^l,\mathbf{b}^l)
            &=\mathrm{Chunk}_l(\hat{\mathbf{e}}^l),\\
        \hat{\mathbf{z}}^L&=\mathcal{B}(\mathbf{e}^L).
    \end{aligned}
    \label{eq:unet_architecture}
\end{equation}
Here, $\mathbf{p}^l\in[0,1]^{N_l}$ contains boundary probabilities and $\mathbf{b}^l\in\{0,1\}^{N_l}$ contains binary selections. The retained states form $\mathbf{e}^{l+1}\in\mathbb{R}^{N_{l+1}\times d_{l+1}}$. The bottleneck $\mathcal{B}$ processes the coarsest sequence. For $l=L-1,\ldots,0$, the decoder restores positions and injects skip features
\begin{equation}
    \begin{aligned}
        \mathbf{z}^l&=\mathrm{Dechunk}_l
        (\hat{\mathbf{z}}^{l+1},\mathbf{p}^l,\mathbf{b}^l)
        +A^l(\hat{\mathbf{e}}^l),\\
        \hat{\mathbf{z}}^l&=D^l(\mathbf{z}^l).
    \end{aligned}
\end{equation}
The projection $A^l$ aligns the skip feature width. The encoder, bottleneck, and decoder use causal Mamba-2 blocks~\cite{dao2024transformers}. Width changes append a shared learnable vector during expansion and retain leading channels during reduction~\cite{slagle2024spacebyte}. The final state $\hat{\mathbf{z}}^0$ has one feature vector per original byte position and gives the representation in Eq.~\eqref{eq:rep_dc}.

\subsection{Dynamic chunking}
At level $l$, two learned projections produce $\mathbf{q}_j^l=W_q\hat{\mathbf{e}}_j^l$ and $\mathbf{u}_j^l=W_u\hat{\mathbf{e}}_j^l$. Following dynamic chunking~\cite{hwang2025dynamic}, the boundary probability is
\begin{equation}
    p_j^l=\frac{1}{2}\left(1-
    \frac{\mathbf{q}_j^l\cdot\mathbf{u}_{j-1}^l}
    {\|\mathbf{q}_j^l\|\|\mathbf{u}_{j-1}^l\|}\right),\qquad p_1^l=1.
\end{equation}
The selection is $b_j^l=\mathbb{I}\{p_j^l\geq0.5\}$, and the next level retains the states at selected positions. These are learned boundaries in the representation sequence. The model is not supplied with plaintext-region annotations or supervised protocol boundaries.

During dechunking, let $v_m^l$ be the boundary probability of retained state $m$. Exponential smoothing gives
\begin{equation}
    \overline{\mathbf{z}}_m^{l+1}
    =v_m^l\hat{\mathbf{z}}_m^{l+1}
    +(1-v_m^l)\overline{\mathbf{z}}_{m-1}^{l+1},
\end{equation}
with $\overline{\mathbf{z}}_1^{l+1}=\hat{\mathbf{z}}_1^{l+1}$. For original position $j$, define $m(j)=\sum_{n=1}^j b_n^l$ and the confidence $c_j^l=b_j^lp_j^l+(1-b_j^l)(1-p_j^l)$. Causal expansion with a straight-through estimator~\cite{bengio2013estimating} is
\begin{equation}
    \widetilde{\mathbf{z}}_j^l=
    \left[c_j^l+\operatorname{stopgradient}(1-c_j^l)\right]
    \overline{\mathbf{z}}_{m(j)}^{l+1}.
\end{equation}
The bracket equals one in the forward pass, while its differentiable component carries the confidence gradient. The expanded states are combined with the skip features before decoder refinement.

\subsection{Prediction and ratio regularization}
A linear head projects $\hat{\mathbf{z}}_i^0$ onto the byte vocabulary to compute $P_{\theta_b}(t_{i+1}\mid t_{\leq i})$ in Eq.~\eqref{eq:dc_loss}. For each transition, the discrete selected fraction and mean boundary probability are
\begin{equation}
    J^l=\frac{1}{N_l}\sum_{j=1}^{N_l}b_j^l,\qquad
    G^l=\frac{1}{N_l}\sum_{j=1}^{N_l}p_j^l.
\end{equation}
For target compression ratio $C^l>1$, the structural regularizer is
\begin{equation}
    \begin{aligned}
    \mathcal{L}_{\mathrm{ratio}}=
    \sum_{l=0}^{L-1}\frac{C^l}{C^l-1}\Bigl[
    &(C^l-1)J^lG^l\\
    &+(1-J^l)(1-G^l)\Bigr].
    \end{aligned}
\end{equation}
It encourages retained fractions near $1/C^l$~\cite{hwang2025dynamic}. Independent pretraining uses next-byte prediction plus this regularizer. Downstream TrafficDCNet training uses classification loss plus $\alpha\mathcal{L}_{\mathrm{ratio}}$, with $\alpha=0.03$ for unfrozen training and $\alpha=0$ for frozen evaluation.

\section{AlignTraffic Objectives}\label{app:align}
AlignTraffic aligns each sample's ByteBERT and TrafficDCNet representations from Eqs.~\eqref{eq:rep_sb} and~\eqref{eq:rep_dc}. MLP heads $g_w$ and $g_b$ project their global features of widths $d_w$ and $d_b$ into normalized embeddings of width $d_{\mathrm{emb}}$.
\begin{equation}
    \begin{aligned}
        \mathbf{h}_w & =
        \frac{g_w(\overline{\mathbf{R}}_w)}
        {\lVert g_w(\overline{\mathbf{R}}_w)\rVert_2}
        \in\mathbb{R}^{d_{\mathrm{emb}}},
        \mathbf{h}_b  =
        \frac{g_b(\overline{\mathbf{R}}_b)}
        {\lVert g_b(\overline{\mathbf{R}}_b)\rVert_2}
        \in\mathbb{R}^{d_{\mathrm{emb}}}.
    \end{aligned}
    \label{eq:rep_proj}
\end{equation}
As shown in Fig.~\ref{fig:align}, fusion encoder $f_m$ applies cross-attention to both token sequences and produces token and global outputs of width $d_f$.
\begin{equation}
    \begin{aligned}
        \mathbf{R}_{wb}
         & =f_m(\mathbf{R}_w,\mathbf{R}_b)
        \in\mathbb{R}^{N_0\times d_f}, \\
        \overline{\mathbf{R}}_{wb}
        =\mathrm{Pool}(\mathbf{R}_{wb})
        \in\mathbb{R}^{d_f}.
    \end{aligned}
    \label{eq:rep_fusion}
\end{equation}

\paragraph{Cross-view contrastive loss}
The contrastive objective aligns the two views by maximizing the similarity of matched pairs and minimizing the similarity to mismatched samples. Inspired by ALBEF~\cite{li2021align} and MoCo~\cite{he2020momentum}, we adopt momentum encoders and feature queues to stabilize training and to enlarge the negative set.
Concretely, the set $\mathcal{I}=\{1,\ldots,B\}$ indexes a mini-batch. For $v\in\{w,b\}$, $\theta_v$ and $\tilde{\theta}_v$ are the online and momentum encoder-head parameters, and $\mu\in[0,1)$ is the momentum coefficient. The momentum branch produces $\tilde{\mathbf{h}}_v^{(i)}$ under $\tilde{\theta}_v\leftarrow\mu\tilde{\theta}_v+(1-\mu)\theta_v$. Each queue $\mathbf{Q}_v\in\mathbb{R}^{K\times d_{\mathrm{emb}}}$ stores $K$ previous momentum embeddings. The matrix $\tilde{\mathbf{H}}_v\in\mathbb{R}^{(B+K)\times d_{\mathrm{emb}}}$ stacks the current momentum keys in sample order followed by $\mathbf{Q}_v$. The one-hot vector $\mathbf{e}_i\in\{0,1\}^{B+K}$ identifies the $i$th current key. For temperature $\tau>0$ and distillation weight $\gamma\in[0,1]$, cross-entropy is $\mathrm{CE}(\mathbf{q},\mathbf{p})=-\sum_{j=1}^{B+K}q_j\log p_j$. For $(v,v')\in\{(w,b),(b,w)\}$, the distributions, soft targets, and symmetric loss are
\begin{equation}
    \begin{aligned}
         & \mathbf{p}_i^{v\rightarrow v'}
        =\operatorname{softmax}\!
        \left(\tilde{\mathbf{H}}_{v'}\mathbf{h}_v^{(i)}/\tau\right),                                       \\
         & \mathbf{q}_i^{v\rightarrow v'}
        =\gamma\operatorname{softmax}\!
        \left(\tilde{\mathbf{H}}_{v'}\tilde{\mathbf{h}}_v^{(i)}/\tau\right) \quad +(1-\gamma)\mathbf{e}_i, \\
         & \mathcal{L}_{\mathrm{C}}
        =\frac{1}{2B}\sum_{i=1}^{B}
        \Bigl[
            \mathrm{CE}(\mathbf{q}_i^{w\rightarrow b},
            \mathbf{p}_i^{w\rightarrow b})+
            \mathrm{CE}(\mathbf{q}_i^{b\rightarrow w},
            \mathbf{p}_i^{b\rightarrow w})
            \Bigr].
    \end{aligned}
    \label{eq:dis}
\end{equation}

\paragraph{Cross-view matching loss}
Cross-view matching trains the fusion branch to discriminate whether a two-view pair originates from the same encrypted traffic.
Specifically, for $B\geq2$, each $i\in\mathcal{I}$ contributes one positive pair and two hard negative pairs, constructed as
\begin{equation}
    \begin{aligned}
         & \Pr(j_i^-=j)
        =\frac{\exp(s_{ij})}
         {\sum_{r\in\mathcal{I}\setminus\{i\}}\exp(s_{ir})}, \\
         & \Pr(k_i^-=k)
        =\frac{\exp(s_{ki})}
         {\sum_{r\in\mathcal{I}\setminus\{i\}}\exp(s_{ri})}, \\
         & \mathcal{S}_{\mathrm{M}}
        =\biguplus_{i=1}^{B}
        \bigl\{(i,i,1),(i,j_i^-,0),(k_i^-,i,0)\bigr\},
        |\mathcal{S}_{\mathrm{M}}|
        =3B,
    \end{aligned}
\end{equation}
where $s_{ij}=\mathbf{h}_w^{(i)\top}\mathbf{h}_b^{(j)}/\tau$ is the cross-view similarity and $j_i^-,k_i^-\in\mathcal{I}\setminus\{i\}$ are the byte-view and word-view negatives selected based on $\Pr$, respectively. The operator $\biguplus$ denotes multiset union, and a tuple $(a,b,y)$ represents $(\mathbf{u}^{(a)},\mathcal{T}^{(b)})$ with label $y\in\{0,1\}$.

For each $(a,b,y)\in\mathcal{S}_{\mathrm{M}}$, the fusion encoder produces $\overline{\mathbf{R}}_{wb}^{(a,b)}=\mathrm{Pool}(f_m(\mathbf{R}_w^{(a)},\mathbf{R}_b^{(b)}))$. The scalar matching head $g_{\mathrm{M}}$ assigns probability $m_{ab}=\sigma\left(g_{\mathrm{M}}(\overline{\mathbf{R}}_{wb}^{(a,b)})\right)$, where $\sigma$ is the sigmoid function. The matching loss is
\begin{equation}
    \begin{aligned}
        \mathcal{L}_{\mathrm{M}}
         & =-\frac{1}{3B}\!
        \sum_{(a,b,y)\in\mathcal{S}_{\mathrm{M}}}
        \bigl[y\log m_{ab}+(1-y)\log(1-m_{ab})\bigr].
    \end{aligned}
\end{equation}

\paragraph{MLM loss}
To maintain contextualized representations during alignment pretraining, we also adopt an auxiliary fused-view MLM objective. In each alignment mini-batch, $\mathcal{M}^{(i)}$ and $\tilde{\mathbf{u}}^{(i)}$ are resampled using the same masking and token-corruption policy as in independent ByteBERT pretraining. Within each sample, the same masked positions are shared across the two views, where the corresponding raw bytes are replaced with samples from $\operatorname{Unif}(\mathcal{X})$ to form $\tilde{\mathcal{T}}^{(i)}$. The head $g_{\mathrm{MLM}}$ maps fused features to logits over $\mathcal{X}$. The target-token probability and fused-view MLM loss are
\begin{equation}
    \begin{aligned}
        \tilde{\mathbf{R}}_{wb}^{(i)}
         & =f_m\!\left(f_w(\tilde{\mathbf{u}}^{(i)}),
        f_b(\tilde{\mathcal{T}}^{(i)})\right),                               \\
        \rho_{ik}
         & =\left[\operatorname{softmax}\!
                \left(g_{\mathrm{MLM}}
                (\tilde{\mathbf{R}}_{wb,k}^{(i)})\right)\right]_{u_k^{(i)}}, \\
        \mathcal{L}_{\mathrm{MLM}}^{wb}
         & =-\frac{1}{N_{\mathrm{mask}}}
        \sum_{i=1}^{B}\sum_{k\in\mathcal{M}^{(i)}}\log\rho_{ik}.
    \end{aligned}
\end{equation}

The complete objective is Eq.~\eqref{eq:loss_align}. Independent ByteBERT pretraining selects valid positions with probability 0.15 and predicts their original values using Eq.~\eqref{eq:byte_mlm}. Alignment pretraining resamples the masked positions and synchronizes them between the branches. The raw-byte branch uses uniformly sampled replacement bytes at those positions, as specified above. This avoids leaving an uncorrupted raw-byte target available to the fusion module at a selected position.

\section{Training Settings}\label{app:hyper}
Experiments use Ubuntu 24.04, CUDA 12.8, and PyTorch 2.8.0. Multi-GPU training uses distributed data parallelism on one node. The following settings describe the reported experiments.

\subsection{Contextual baselines and ByteBERT}
Our ET-BERT implementation uses HuggingFace BERT while retaining the original vocabulary and model design. Pretraining uses 100 epochs, batch size 128, learning rate $2\times10^{-5}$, and warmup ratio 0.1. Downstream training uses AdamW for 50 epochs, batch size 128, and dropout 0.5. The unfrozen learning rates are $6\times10^{-5}$ for flow-level and $2\times10^{-5}$ for packet-level classification. Frozen classifier training uses $2\times10^{-4}$. ByteBERT uses the same backbone configuration and training settings, with single-byte tokens and masking probability 0.15.

YaTC uses AdamW with linear learning-rate scaling, base pretraining rate $10^{-3}$, batch size 128, and 150,000 pretraining steps. The patch masking ratio is 0.75. Fine-tuning uses 200 epochs, base learning rate $2\times10^{-3}$, and batch size 128. NetMamba follows its official implementation with the same base pretraining rate, batch size, and number of steps. Its stride masking ratio is 0.9, and its downstream learning rate is $2\times10^{-3}$~\cite{wang2024netmamba}.

\subsection{TrafficDCNet and AlignTraffic}
TrafficDCNet uses three hierarchical levels with widths $\{512,768,1024\}$. Two transitions each target a compression ratio of 3, giving cumulative ratios $\{1,3,9\}$. Independent pretraining uses AdamW for 1000 epochs with learning rate $8\times10^{-4}$, $\beta=(0.9,0.95)$, weight decay 0.01, batch size 128, and a warmup, hold, and decay schedule. The ratio-loss weight is 0.03. Downstream training uses 50 epochs, learning rate $3\times10^{-4}$, and batch size 128.

AlignTraffic uses a 12-layer BERT-base pathway. Its first six layers form the word-view encoder $f_w$, and the remaining six layers become the cross-attention fusion module $f_m$. The byte pathway reuses TrafficDCNet. Alignment training uses AdamW for 50 epochs, including 10 warmup epochs, with batch size 128. The initial, minimum, and warmup learning rates are $5\times10^{-5}$, $10^{-5}$, and $5\times10^{-6}$, respectively. Weight decay is 0.02, followed by a cosine learning-rate schedule. We set $\tau=0.07$, queue size $K=65{,}536$, momentum $\mu=0.995$, distillation weight $\gamma=0.4$, MLM weight $\eta=1$, and ratio-loss weight $\alpha=0.03$. Downstream fine-tuning uses AdamW for 50 epochs, batch size 128, and learning rate $10^{-4}$.

\subsection{Protocol-aware and shallow baselines}
Pcap-Encoder starts from the authors' released T5-AE+Q\&A checkpoint and uses their two-layer MLP classifier. Downstream AdamW learning rates are $3\times10^{-4}$ for frozen and $10^{-4}$ for unfrozen training. Flow-level classification trains a packet classifier on the first five packets and uses majority voting for the flow label~\cite{10.1145/3718958.3750498}. The flow-level frozen results are not reported.

The shallow baseline uses AutoGluon TabularPredictor to select among LightGBM, XGBoost, and Random Forest~\cite{erickson2020autogluon}. LightGBM and XGBoost use its default tabular configurations. Random Forest candidates use Gini and entropy splitting criteria. The feature groups are described below.

\section{Shallow Feature Groups}\label{app:shallow}
The packet-level features describe IP and TCP behavior, generic TLS records, and visible ServerHello information. IP features include length, flags, fragmentation, checksum, protocol, header length, type of service, TTL, version, and identification. TCP features include sequence and acknowledgment numbers, flags, checksum, window, header length, urgent pointer, reserved fields, timestamps, and selective acknowledgment options. Original 5-tuple fields are suppressed in the masked comparison.

TLS features include payload length, TLS presence, the first record's type, version, and length, record count and length summaries, and application-data, handshake, alert, and change-cipher-spec statistics. When a complete visible ServerHello is available, the features include its selected version, cipher suite, extension count, supported version, and key-share group. ALPN is included when visible in the captured protocol version.

Flow features expand the packet groups across the first five packet positions, add four inter-packet intervals, and aggregate TLS information across those packets. The aggregates include ServerHello recovery and attributes, session-ID length, directional counts, and directional byte statistics. This baseline therefore has explicit access to parsed protocol information, including fields within the retained payload.

\section{Industrial Evaluation Details}\label{app:industrial}

Script-driven applications generate the industrial traffic across multiple terminal devices and network-quality conditions. Traffic is captured at an intermediate network device. The 165 application classes include voice, video, gaming, social media, and AIGC services, with ground-truth labels derived from plaintext-visible information. Preprocessing retains payloads from flows with at least eight packets and removes packet headers. The training set is relatively balanced, while the test set follows an imbalanced deployment distribution.

We evaluate ET-BERT, YaTC, AlignTraffic, and the deployed industrial baseline in the unfrozen setting. The payload-only inputs omit the header fields used by Pcap-Encoder and our shallow baseline. The collaborator restricts release of the dataset because it contains private traffic information.

\IEEEtriggeratref{29}
\bibliographystyle{IEEEtran}
\bibliography{references}

\end{document}